\documentclass{article}
\usepackage{latexsym}
\usepackage{graphicx}
\usepackage{float}
\usepackage{bm}
\usepackage{subfigure}
\usepackage{caption}
\usepackage{makecell,rotating}
\usepackage{amssymb}

\begin{document}
\begin{center}
\textbf{Feasibility study on the least square method for fitting non-Gaussian noise data}
\end{center}

\begin{center}
Wei Xu, Wen Chen, Yingjie Liang
\end{center}

\begin{center}
Institute of Soft Matter Mechanics, College
of Mechanics and Materials, Hohai University, Nanjing, China
\end{center}

\begin{center}
State Key Laboratory of Hydrology-Water Resources and
Hydraulic Engineering
\end{center}

\noindent\textbf{Corresponding authors}:\\

\noindent Wen Chen, Email address: \underline {chenwen@hhu.edu.cn}\\

\noindent Yingjie Liang, Email address: \underline {liangyj@hhu.edu.cn}\\

\noindent\textbf{\textit{Abstract:}} This study is to investigate the
feasibility of least square method in fitting non-Gaussian noise
data. We add different levels of the two typical non-Gaussian
noises, L\'{e}vy and stretched Gaussian noises, to exact value of
the selected functions including linear equations, polynomial and
exponential equations, and the maximum absolute and the mean square
errors are calculated for the different cases. L\'{e}vy and
stretched Gaussian distributions have many applications in
fractional and fractal calculus. It is observed that the
non-Gaussian noises are less accurately fitted than the Gaussian
noise, but the stretched Gaussian cases appear to perform better
than the L\'{e}vy noise cases. It is stressed that the least-squares
method is inapplicable to the non-Gaussian noise cases when the
noise level is larger than 5{\%}.\\


\noindent\textbf{1. Introduction}\\

Non-Gaussian noise is universal in nature and engineering.$^{1-3}$ In recent
decades, non-Gaussian noise has widely been studied, especially in signal
detection and processing,$^{\, 4-5}$ theoretical model analysis,$^{\, 6}$
and error statistics.$^{\, 7}$ It is known that the Gaussian distribution is
the mathematical precondition to use the least square method.$^{4\,
}$However, it is often directly used to process such non-Gaussian noise
data, which may give wrong estimation.$^{\, 8}$ Thus, this study is to
quantitatively examine the applicability of the least square method to
analyze non-Gaussian noise data.

Generally, non-Gaussian noise has detrimental influence on the
stability of power system, and also can stimulate systems to
generate ordered patterns.$^{9-10}$ To our best knowledge, L\'{e}vy
and stretched Gaussian noises are two kinds of typical non-Gaussian
noise, which are frequently used in fractional and fractal
calculus.$^{11-12}$ L\'{e}vy noise has been observed in many
complex systems, such as turbulent fluid flows,$^{13\, }$signal
processing,$^{14}$ financial times series,$^{15-16}$ neural
networks.$^{17\, }$We also note that the parameters estimation for
stochastic differential equations driven by small L\'{e}vy noise
were investigated.$^{18}$ Compared to the L\'{e}vy noise the
stretched Gaussian noise is less studied, but its corresponding
stretched Gaussian distribution has been explored,$^{19}$ such as in
the motion of flagellate protozoa,$^{20\, }$SoL interchange
turbulence simulation,$^{21\, }$anomalous diffusion of particles
with external force,$^{22}$ not mentioned too much. It also should
be pointed that processing of non-sinusoidal signals or sound
textures has become an important research topic,$^{23}$ and the
derived algorithms significantly improve the perceptual quality of
stretched noise signals.$^{19\, }$

It is well known that the least square method is a standard regression
approach to approximate the solutions of over determined systems, which is
most frequently used in data fitting and estimation.$^{24}$ The core concept
of the least square method is to identify the best match for the system by
minimizing the square error.$^{25}$ Supposed that the data points are
${(x_1,\,y_1)},\,{(x_2,\,y_2)}\,{(x_3,\,y_3)},...,{(x_{200},\,y_{200})},$ where $x$ represents the independent
variable and $y$ is the dependent variable. The fitting error $d$
characterizes the distance between $y$ and the estimated curve$f\left( x
\right)$, i.e., $d_{1} =y_{1} -f\left( {x_{1} } \right),\;...,d_{n} =y_{n}
-f\left( {x_{\mbox{n}} } \right)$. The best fitting curve is to minimize the
square error $\delta =\;d_{1}^{2}+d_{2}^{2}+...+d_{\mbox{n}}
^{2}=\sum\limits_{i=1}^n {\left[ {y_{i} -f\left( {x_{i} } \right)} \right]}
^{2}$, where the errors $d_{i} $ are usually modeled by Gaussian
distribution.$^{26}$

Field data are often polluted by noise$^{27\, }$and the Gaussian noise is
the classical one, whose probability density function obeys Gaussian
distribution. We have mentioned above that several types of noise data obey
non-Gaussian distribution.$^{28}$ To examine the feasibility of the least
square method in fitting non-Gaussian noise data, we generate the
non-Gaussian random numbers as the noise, and then add different levels of
the noise to the exact values of the selected functions including linear
equations, polynomial and exponential equations as the observed values. By
using the least square method, the maximum absolute and mean square errors
are calculated and compared in the Gaussian and non-Gaussian applications.

The rest of the paper is organized as follows. In Section 2, we
introduce the Gaussian distribution, L\'{e}vy distribution,
stretched Gaussian distribution, and the methods we use to analyze
the noise data. In Section 3, we give the results and discussion.
Finally, a brief summary is provided.\\

\noindent\textbf{2. Theory and methods}\\

\noindent\textbf{\textit{2.1 Gaussian distribution }}\\

Gaussian distribution is also called as normal distribution, which is often
encountered in mathematics, physics and engineering. The probability density
function of Gaussian distribution is:
\begin{equation}
\label{eq1}
f\left( {x;\mu ,\sigma } \right)=\frac{1}{\sqrt {2\pi \sigma^{2}} }\exp
\left( {-\frac{\left( {x-\mu } \right)^{2}}{2\sigma^{2}}} \right),
\end{equation}
where $\mu $ and $\sigma $ respectively represent mean and standard
deviation. When $\mu =0$ and $\sigma =1$, it degenerates into the
standard Gaussian distribution. Figure1 gives four different cases
of Gaussian density function.\\
\begin{figure}[htbp]
\centering
\includegraphics[width=2.5in,height=2in]{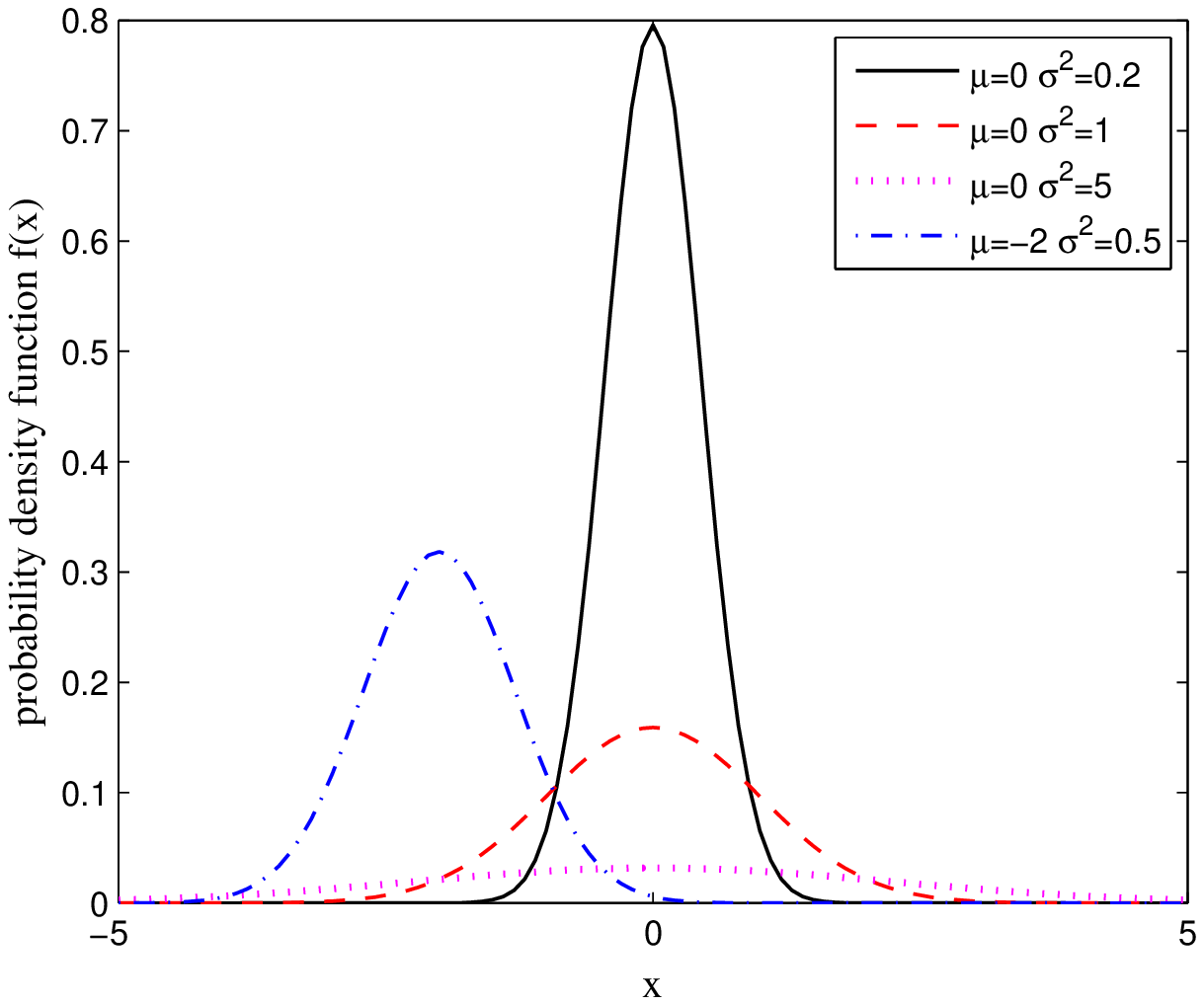}

\small{{\textbf{Figure 1. The probability density functions of Gaussian distribution.}}}
\label{fig1}
\end{figure}

\noindent\textbf{\textit{2.2 L\'{e}vy distribution}}\\

L\'{e}vy distribution, named after Paul L\'{e}vy, is a rich class of
probability distributions. The Gaussian and Cauchy distributions are its
special cases. It is usually defined by its characteristic
function $\phi_{\alpha ,\beta } \left( {k;\mu ,\sigma } \right)$.$^{29}$
\begin{equation}
\label{eq2}
\phi_{\alpha ,\beta } \left( {k;\mu ,\sigma } \right)=F\left\{ {f_{\alpha
,\beta } \left( {k;\mu ,\sigma } \right)} \right\}=\exp \left[ {i\mu
k-\sigma^{\alpha }\left| k \right|^{\alpha }\left( {1-i\beta
\frac{k}{\left| k \right|}\omega \left( {k,\alpha } \right)} \right)}
\right],
\end{equation}
where
\begin{equation}
\label{eq3}
\omega \left( {k,\alpha } \right)=\left\{ {\begin{array}{l}
 \tan \frac{\pi \alpha }{2}\;\;\;\;\;\;\;\;\;\;\;\left( {\alpha \ne
1,0<\alpha <2} \right) \\
 -\frac{2}{\pi }\ln \left| k \right|\;\;\;\;\;\;\;\;\;\left( {\alpha =1}
\right) \\
 \end{array}} \right.\quad ,
\end{equation}
stability index $0<\alpha \le 2$, skewness parameter $-1\le \beta \le 1$,
scale parameter $\sigma >0$, and location parameter $\mu \in \Re $. $\alpha
$ and $\beta $ respectively determine the properties of asymptotic decay and
symmetry. The standard L\'{e}vy distribution can be obtained by the
following transformation.
\begin{equation}
\label{eq4}
p_{\alpha ,\beta } (k;\mu ,\sigma )=\frac{1}{\sigma }p_{\alpha ,\beta }
(\frac{x-\mu }{\sigma };0,1).
\end{equation}

When$\beta =1,\;\mu =0$, the probability density function of the L\'{e}vy
distribution is stated as:
\begin{equation}
\label{eq5}
f\left( {x;\mu ,c} \right)=\sqrt {\frac{c}{2\pi }} \left(
{-\frac{e^{-\frac{c}{2x}}}{x^{1.5}}} \right),
\end{equation}
where $x\ge 0$, $\mu $ is the location parameter and $c$ is the
scale parameter. Different cases of Eq.(5) are illustrated
in Figure 2.\\

\begin{figure}[htbp]
\centering
\includegraphics[width=2.5in,height=2in]{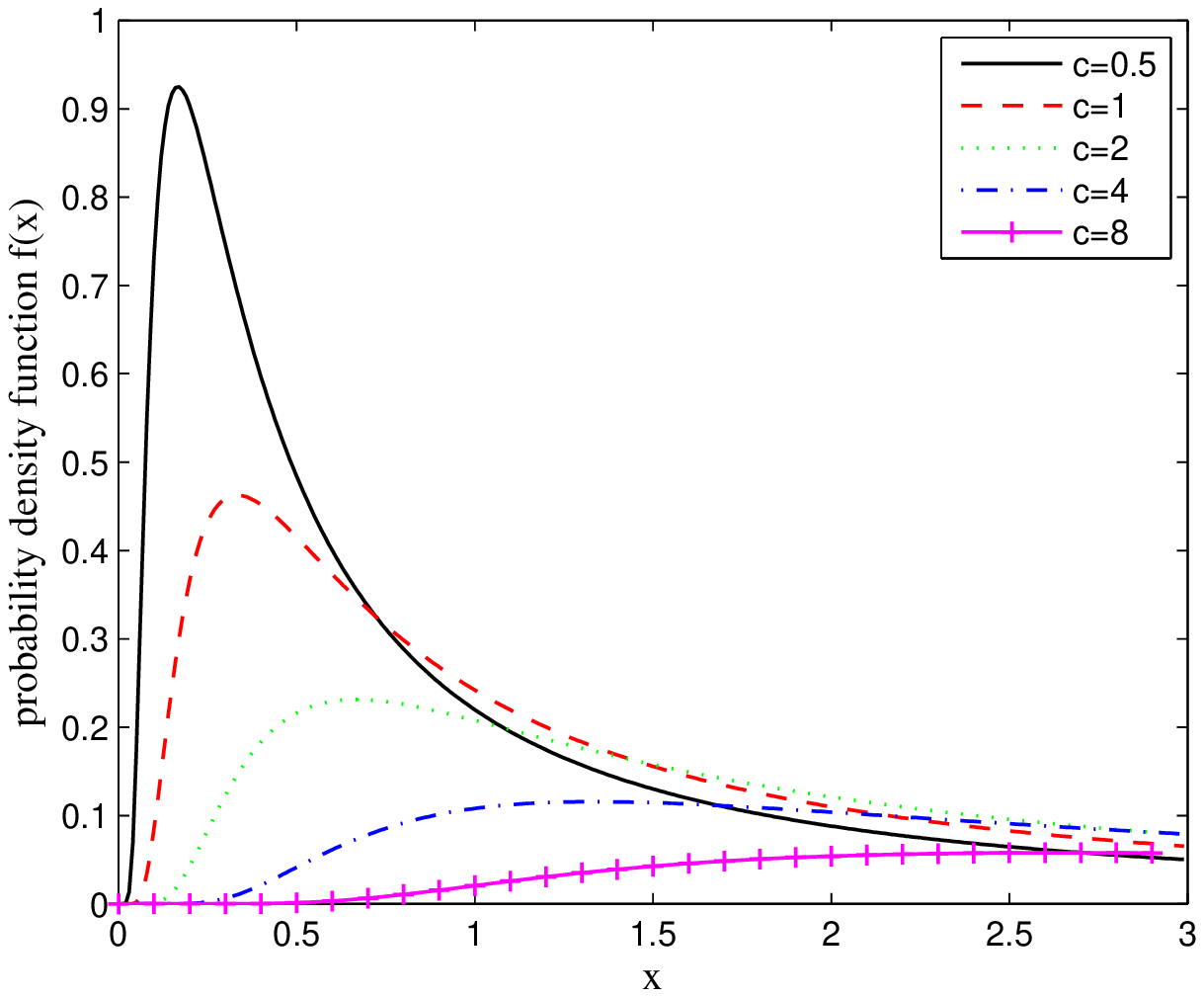}

\small{{\textbf{Figure 2. The probability density functions of L\'{e}vy distribution.}}}
\label{fig2}
\end{figure}

\noindent\textbf{\textit{2.3 Stretched Gaussian distribution}}\\

The stretched Gaussian distribution has widely been used to describe
anomalous diffusion and turbulence, especially in the porous media with
fractal structure.$^{30}$ The solution to the fractal derivative equation in
characterizing the fractal media has the form of stretched Gaussian
distribution,$^{31\, }$whose probability density function is defined as:
\begin{equation}
\label{eq6}
f_{\beta } \left( x \right)=\frac{\beta }{2^{1+1/\beta }\Gamma \left(
{1/\beta } \right)\sigma }\exp \left( {-\frac{1}{2}\left| {\frac{x-a}{\sigma
}} \right|^{\beta }} \right)\;,\;\;\;\;\;\left( {-\infty <x<\infty ,\;\beta
>0} \right).
\end{equation}
where $\beta $ is the stretched exponent. When$\beta =2$ and $a=0$,
it becomes to the standard Gaussian distribution. Figure 3 shows
three cases of stretched Gaussian density function.
\begin{figure}[htbp]
\centering
\includegraphics[width=2.5in,height=2in]{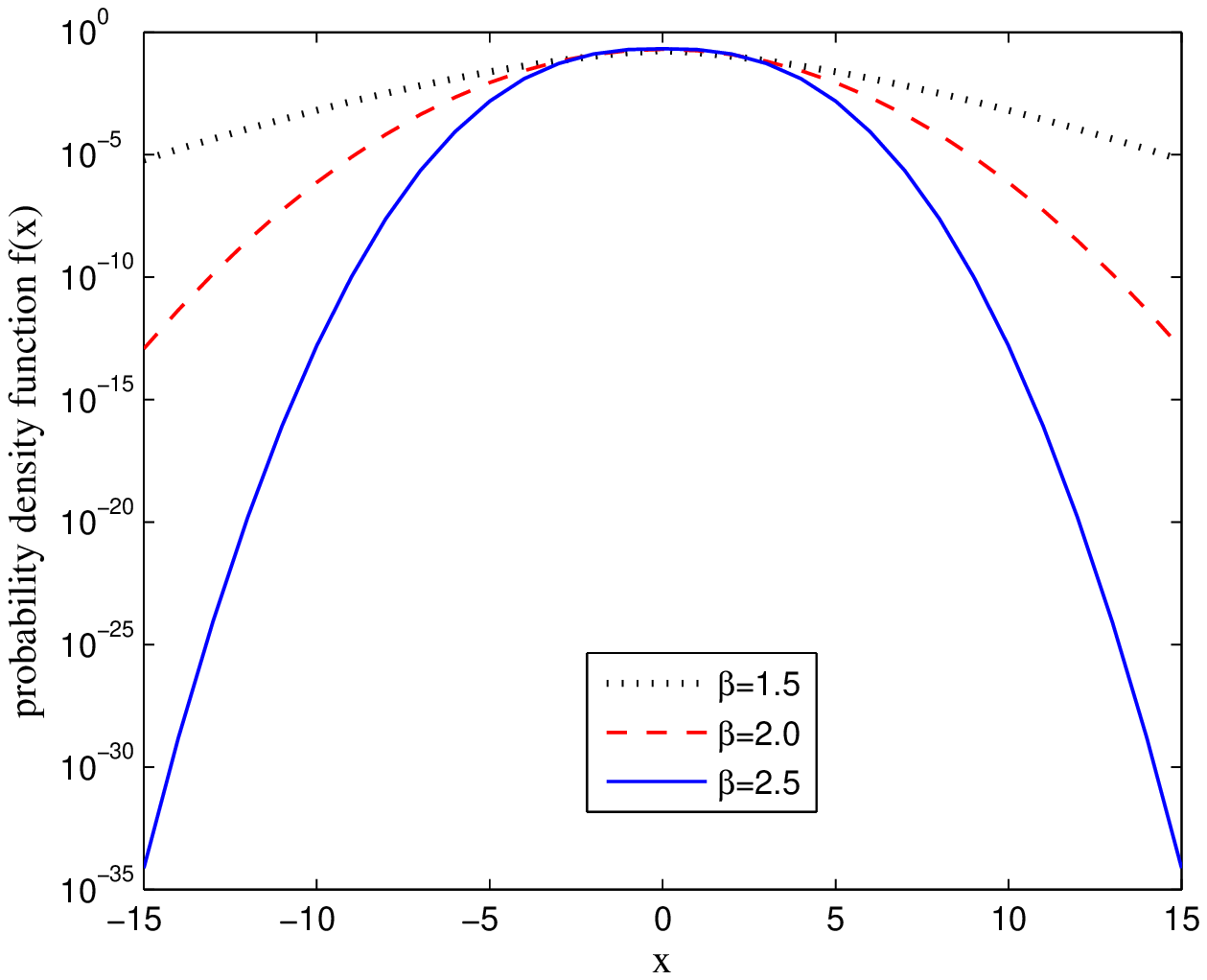}

\small{{\textbf{Figure 3. The probability density functions of stretched Gaussian
distribution.}}}
\label{fig3}
\end{figure}

\noindent\textbf{\textit{2.4 Generation of noise data}}\\

In this study, the noise data are obtained based on the above mentioned
Gaussian and non-Gaussian random variables, which can be generated by using
the inverse function method$^{32\, }$and the selection method.$^{33}$
Specifically, Chambers, Mallows, Stuck proposed the CMS method in L\'{e}vy
random variables simulation,$^{\, 34}$ which is the fastest and most
accuracy method. By using the CMS method, some variables need to be defined
first.
\[
V=\pi \left( {U_{1} -\frac{1}{2}} \right),
\]
\[
W=-\ln U_{2} ,
\]
\[
L=\left\{ {\left. {1+\beta^{2}\tan^{2}\left( {\frac{\pi \alpha }{2}}
\right)} \right\} } \right.^{1/2\alpha },
\]
\[
\theta_{0} =\frac{1}{\alpha }\arctan \left( {\beta \tan \left( {\frac{\pi
\alpha }{2}} \right)} \right),
\]
where $U_{1} $ and $U_{2} $are two independent uniform distribution on
interval $\left( {0,1} \right)$.

When $\alpha \ne 1$, the {L\'{e}vy} random number is
\begin{equation}
\label{eq7}
X=L\frac{\sin \left\{ {\left. {\alpha \left( {V+\theta_{0} } \right)}
\right\} } \right.}{\left\{ {\cos \left( V \right)} \right\}^{1/\alpha
}}.\left[ {\frac{\cos \left( {V-\alpha \left( {V+\theta_{0} } \right)}
\right)}{W}} \right]^{\left( {1-\alpha } \right)/\alpha }.
\end{equation}

When $\alpha =1$,
\begin{equation}
\label{eq8}
X=\frac{2}{\pi }\left\{ {\left( {\frac{\pi }{2}+\beta V} \right)\tan V-\beta
\ln \left( {\frac{\frac{\pi }{2W}\cos V}{\frac{\pi }{2}+\beta V}} \right)}
\right\}.
\end{equation}

The general L\'{e}vy random numbers can be obtained based on some known
properties.$^{35}$

For the stretched Gaussian distribution, we use the acceptance
rejection method to generate its random numbers.$^{\, 33}$\\

\noindent\textbf{\textit{2.5 Methods}}\\

Both linear and nonlinear functions estimation are considered by using the
least square method, in which the model function $f:\;R^{m}\to R$ is
estimated as
\begin{equation}
\label{eq9}
y_{i} =f\left( {x_{i} } \right)+\varepsilon_{i} \;\left( {i=1,2,...,n}
\right),
\end{equation}
where $n$ is the number of observations, $y_{i} \in R$ the response
variable, $x_{i} \in R^{m}$ the explanatory variable, the noise
\begin{equation}
\label{eq9}
\varepsilon_{i} \;\mbox{=}\;rand\times a\mbox{\% \, }\left(
{a=1,\;5,\;10,\;15,\;20} \right).
\end{equation}

In this study, we consider the case$n=200$, then the observed values $y_{1}
,\;y_{2} ,...,\;y_{200} $ can be constructed by adding the values of the
random numbers to the exact values of the selected functions including
linear equations, polynomial and exponential equations, finally the maximum
absolute error and the mean square error are calculated for the above
different cases in conjunction with the least square method.

We give the following abbreviations in noise date processing for
convenience:\\
\noindent{FA: Gaussian noise least square error fitting, $\mu =5,\;\sigma =0.5.$}\\
\noindent{FB: L\'{e}vy noise least square error fitting, $\alpha =1.8,\;\beta =0,\;\mu
=1,\;\sigma =0.$}\\
\noindent{FC: Stretched Gaussian noise least square error fitting,$\beta
=2.5,\;a=1,\;\sigma =3.$}\\
\noindent{Rerr1: Maximum absolute error: $\max \left| {F\left( {x_{i} }
\right)-f\left( {x_{i} } \right)} \right|$.}\\
\noindent{Rerr2: Mean square error: $\sqrt {\frac{\sum\limits_{i=1}^n {\left.
{\left\{ {F\left( {x_{i} } \right)-f\left( {x_{i} } \right)}
\right.} \right\}^{2}} }{n}} $.}\\

\noindent\noindent\textbf{3. Results and discussion}\\

In this section, we apply the least square method to fit various
noise-polluted data by adding different levels of Gaussian and
non-Gaussian noise to exact values of the selected functions
including linear equations, polynomial and exponential equations,
and give a brief discussion.\\

a) The simplest typical model is the linear function.
\begin{equation}
\label{eq12}
f\left( x\right)=ax+b,\;\;\left( {a\ne 0} \right).
\end{equation}

Here we select the following case as example:
\begin{equation}
\label{eq1}
f\left( x \right)=5x.
\end{equation}

Tables 1 to 5 give the estimated parameters and the errors for five
different levels of noise in the linear case. We can observe that the
Gaussian noise fitting data maximum absolute error is in the range of
$\left( {0.0005,\;0.0177} \right)$, the mean square error is in the range of
$\left( {6.8544e-15,\;1.5782e-8} \right)$. The maximum absolute and the mean
square errors of Gaussian noise are the least and L\'{e}vy distribution
noise are the largest, with the relationship expressed as:
\begin{equation}
\label{eq12}
Rerr1\left( {FA} \right)\mbox{\, }<\mbox{\, }Rerr1\left( {FC}
\right)\mbox{\, }<\mbox{\, }Rerr1\left( {FB} \right),
\end{equation}
\begin{equation}
\label{eq13}
Rerr2\left( {FA} \right)\mbox{\, }<\mbox{\, }Rerr2\left( {FC}
\right)\mbox{\, }<\mbox{\, }Rerr2\left( {FB} \right).
\end{equation}

The corresponding fitting curves are depicted in Figures 4-8. We can find
that the results of Gaussian noise fitting have the best accuracy, and the
stretched Gaussian noise fitting curves are closer to those of the Gaussian
noise compared with the results of L\'{e}vy noise data fitting.\\

\noindent Table 1. The estimated results for 1{\%} noise in the linear case.
\begin{center}
\begin{tabular}{|l|p{79pt}|p{79pt}|l|l|}
\hline
\textbf{\textit{function}}$_{\mathbf{}}$&
$a_{\mathbf{}}$&
$b_{\mathbf{}}$&
$_{\mathbf{Rerr1}}$&
$_{\mathbf{Rerr2}}$ \\
\hline
$_{\mathbf{f}}$&
$_{5}$&
$_{0}$&
$_{0}$&
$_{0}$ \\
\hline
$_{\mathbf{FA}}$&
$_{4.9901}$&
$_{0.0005}$&
$_{0.0005}$&
$_{6.8544e-15}$ \\
\hline
$_{\mathbf{FB}}$&
$_{4.9973}$&
$_{-0.0073}$&
$_{0.0076}$&
$_{3.0886e-9}$ \\
\hline
$_{\mathbf{FC}}$&
$_{5.0087}$&
$_{-0.0011}$&
$_{0.0011}$&
$_{2.9304e-13}$ \\
\hline
\end{tabular}
\label{tab1}
\end{center}
\begin{figure}[H]
\centering
\includegraphics[width=2in,height=1.6in]{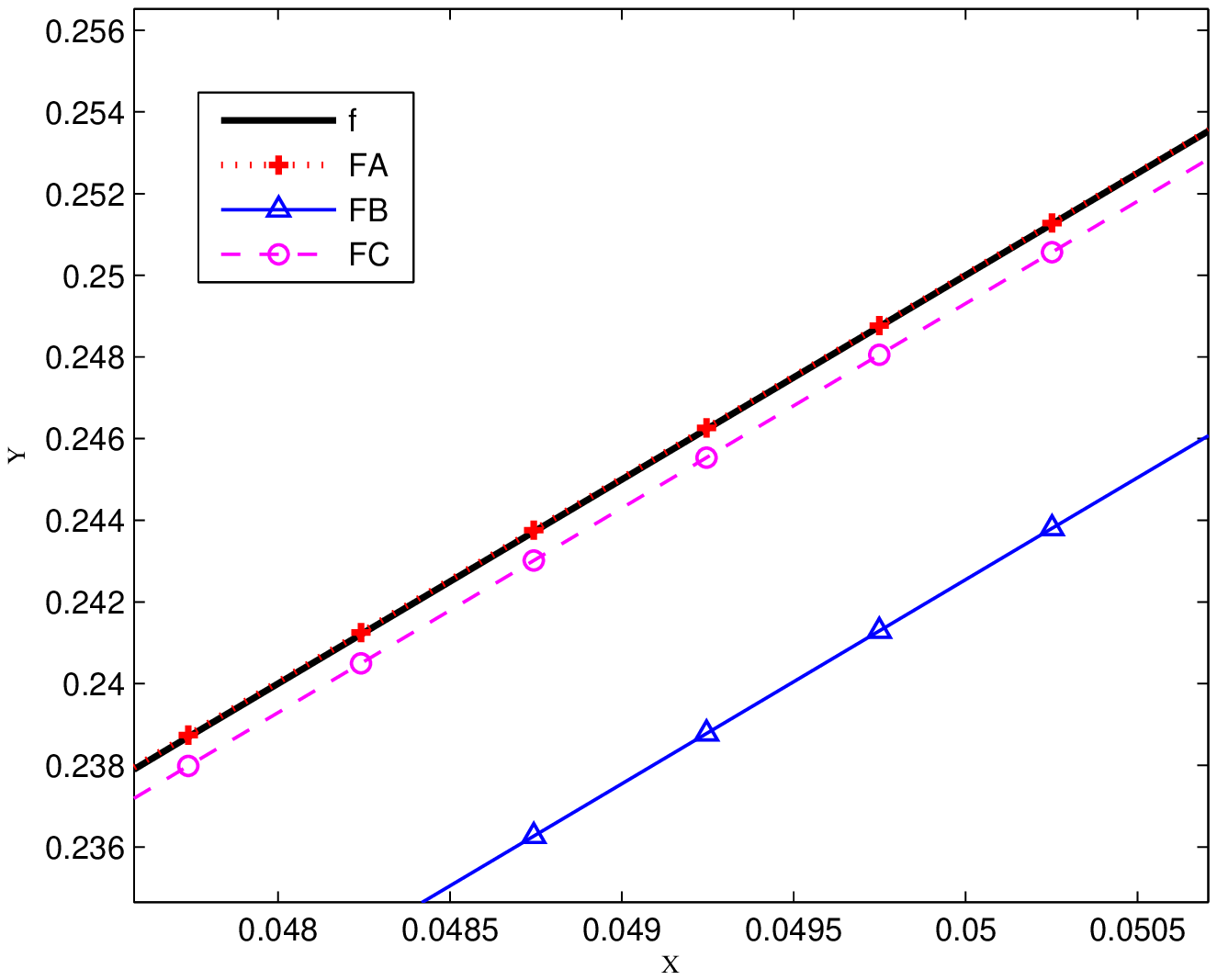}

\small{{\textbf{Figure 4. Linear model fitting to 1{\%} noise.}}}
\label{fig4}
\end{figure}
\noindent{Table 2. }The estimated results for 5{\%} noise in the linear case.
\begin{center}
\begin{tabular}{|l|p{79pt}|p{79pt}|l|l|}
\hline
\textbf{\textit{function}}$_{\mathbf{}}$&
$a_{\mathbf{}}$&
$b_{\mathbf{}}$&
$_{\mathbf{Rerr1}}$&
$_{\mathbf{Rerr2}}$ \\
\hline
$_{\mathbf{f}}$&
$_{5}$&
$_{0}$&
$_{0}$&
$_{0}$ \\
\hline
$_{\mathbf{FA}}$&
$_{5.0195}$&
$_{-0.0009}$&
$_{0.0010}$&
$_{1.0324e-13}$ \\
\hline
$_{\mathbf{FB}}$&
$_{5.0138}$&
$_{0.0102}$&
$_{0.0116}$&
$_{1.4090e-8}$ \\
\hline
$_{\mathbf{FC}}$&
$_{5.0635}$&
$_{-0.0119}$&
$_{0.0109}$&
$_{6.4216e-9}$ \\
\hline
\end{tabular}
\label{tab2}
\end{center}
\begin{figure}[H]
\centering
\includegraphics[width=2in,height=1.6in]{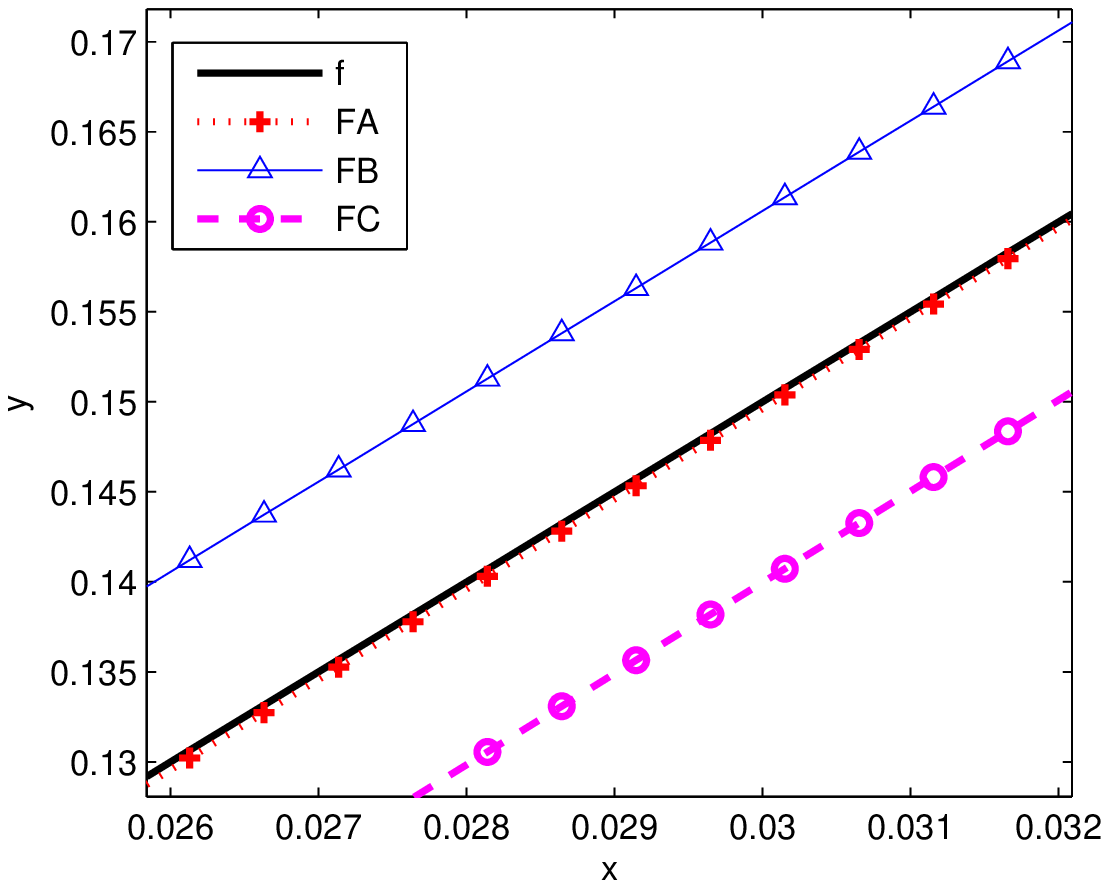}

\small{{\textbf{Figure 5. Linear model fitting to 5{\%} noise.}}}
\label{fig5}
\end{figure}
\noindent{Table 3. }The estimated results for 10{\%} noise in the linear case.
\begin{center}
\begin{tabular}{|l|p{79pt}|p{79pt}|l|l|}
\hline
\textbf{\textit{function}}$_{\mathbf{}}$&
$a_{\mathbf{}}$&
$b_{\mathbf{}}$&
$_{\mathbf{Rerr1}}$&
$_{\mathbf{Rerr2}}$ \\
\hline
$_{\mathbf{f}}$&
$_{5}$&
$_{0}$&
$_{0}$&
$_{0}$ \\
\hline
$_{\mathbf{FA}}$&
$_{5.0504}$&
$_{-0.0032}$&
$_{0.0032}$&
$_{6.5200e-12}$ \\
\hline
$_{\mathbf{FB}}$&
$_{4.9793}$&
$_{0.0104}$&
$_{0.0104}$&
$_{7.7237e-9}$ \\
\hline
$_{\mathbf{FC}}$&
$_{5.1412}$&
$_{-0.0181}$&
$_{0.0101}$&
$_{1.9163e-8}$ \\
\hline
\end{tabular}
\label{tab3}
\end{center}
\begin{figure}[H]
\centering
{\includegraphics[width=2in,height=1.6in]{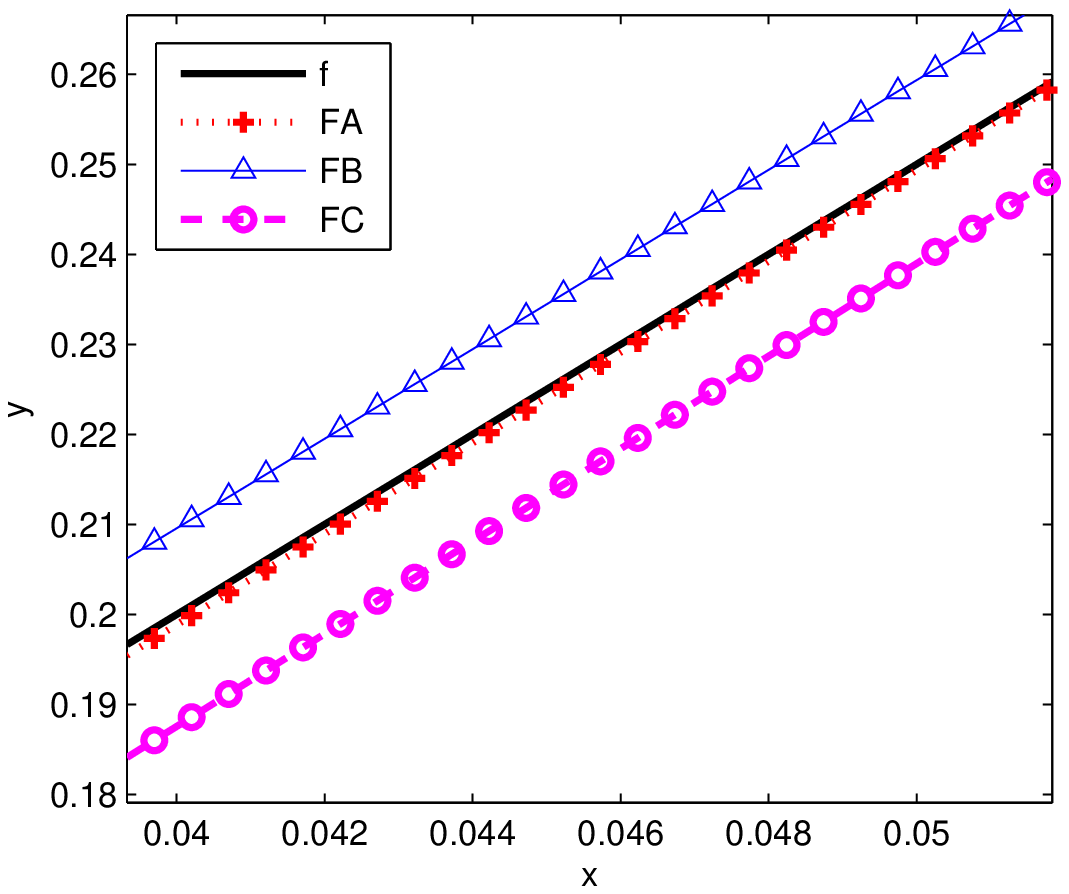}}

\small{{\textbf{Figure 6. Linear model fitting to 10{\%} noise.}}}
\label{fig6}
\end{figure}
\noindent{Table 4. }The estimated results for 15{\%} noise in the linear case.
\begin{center}
\begin{tabular}{|l|p{78pt}|p{78pt}|l|l|}
\hline
{{function}}$_{\mathbf{}}$&
$a_{\mathbf{}}$&
$b_{\mathbf{}}$&
$_{\mathbf{Rerr1}}$&
$_{\mathbf{Rerr2}}$ \\
\hline
$_{\mathbf{f}}$&
$_{5}$&
$_{0}$&
$_{0}$&
$_{0}$ \\
\hline
$_{\mathbf{FA}}$&
$_{4.8624}$&
$_{0.0064}$&
$_{0.0074}$&
$_{2.6201e-10}$ \\
\hline
$_{\mathbf{FB}}$&
$_{4.9193}$&
$_{0.0482}$&
$_{0.0482}$&
$_{3.8108e-6}$ \\
\hline
$_{\mathbf{FC}}$&
$_{5.1177}$&
$_{-0.0322}$&
$_{0.0322}$&
$_{4.9674e-7}$ \\
\hline
\end{tabular}
\label{tab4}
\end{center}
\begin{figure}[H]
\centering
{\includegraphics[width=2in,height=1.6in]{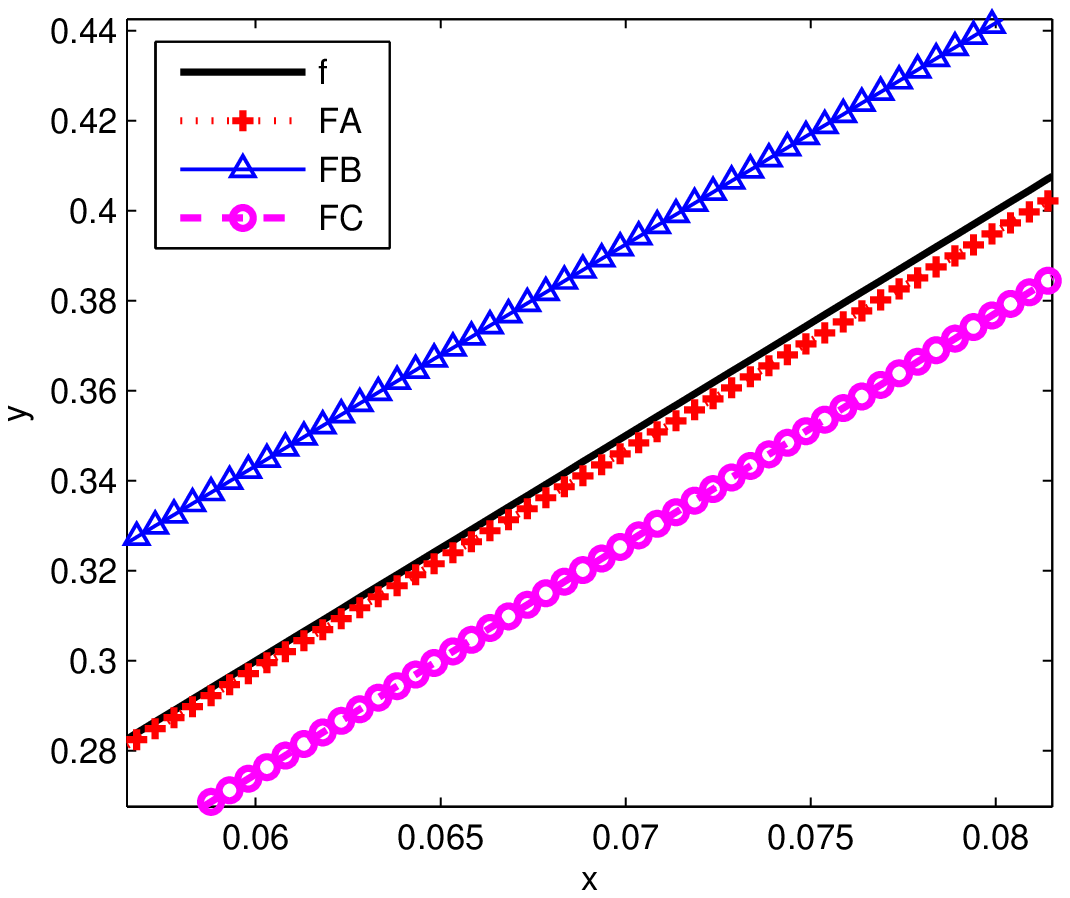}}

\small{\textbf{Figure 7. Linear model fitting to 15{\%} noise.}}
\label{fig7}
\end{figure}

\noindent{Table 5. }The estimated results for 20{\%} noise in the linear case.
\begin{center}
\begin{tabular}{|l|p{70pt}|p{70pt}|l|l|}
\hline
\textbf{\textit{function}}$_{\mathbf{}}$&
$a_{\mathbf{}}$&
$b_{\mathbf{}}$&
$_{\mathbf{Rerr1}}$&
$_{\mathbf{Rerr2}}$ \\
\hline
$_{\mathbf{f}}$&
$_{5}$&
$_{0}$&
$_{0}$&
$_{0}$ \\
\hline
$_{\mathbf{FA}}$&
$_{4.8533}$&
$_{0.0177}$&
$_{0.0177}$&
$_{1.5782e-8}$ \\
\hline
$_{\mathbf{FB}}$&
$_{5.0199}$&
$_{0.1645}$&
$_{0.1665}$&
$_{7.5012e-4}$ \\
\hline
$_{\mathbf{FC}}$&
$_{5.1050}$&
$_{-0.0355}$&
$_{0.0355}$&
$_{8.5735e-7}$ \\
\hline
\end{tabular}
\label{tab5}
\end{center}
\begin{figure}[H]
\centering
{\includegraphics[width=2in,height=1.6in]{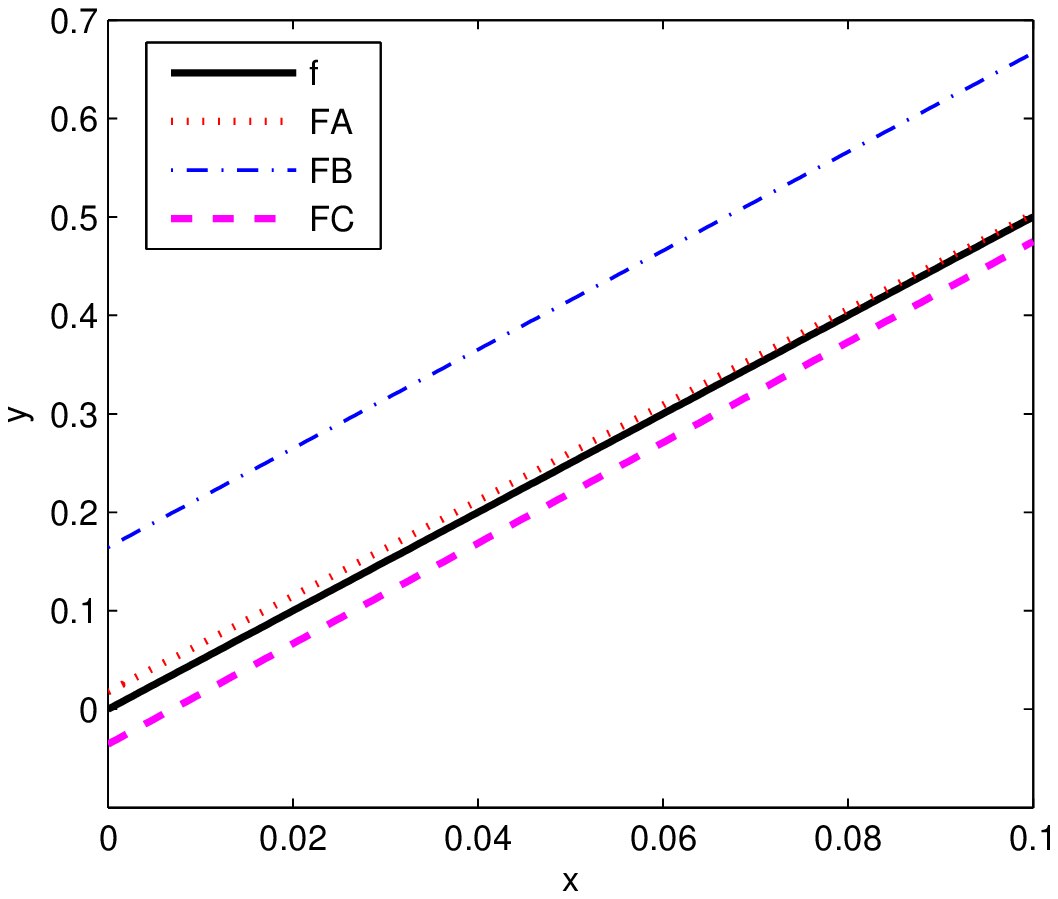}}

\small{\textbf{Figure 8. Linear model fitting to 20{\%} noise.}}
\label{fig8}
\end{figure}

b) A polynomial can be constructed by means of addition, multiplication and
exponentiation to a non-negative power, which is usually written as the
following form with a single variable $x$,
\begin{equation}
\label{eq15}
f\left( x \right)=a_{n} x^{n}+a_{n-1} x^{n-1}+...+a_{2} x^{2}+a_{1} x+a_{0}
,
\end{equation}
where\,$a_{0} ,\;a_{1} ,\;...,\;a_{n-1} ,\;a_{n} $are constants. We select
three parameters polynomial function.
\begin{equation}
\label{eq16}
F\left( x \right)=ax^{2}+bx+c\;\left( {a\ne 0} \right).
\end{equation}

Here the following case is used as example:
\begin{equation}
\label{eq17}
y=4x^{2}+3x+2\;\left( {a\ne 0} \right).
\end{equation}

Tables 6 to10 give the estimated parameters and the errors for five
different levels of noise in the polynomial case. The corresponding fitting
curves are depicted in Figures 9-13. Gaussian noise fitting maximum absolute
error is in the range of $\left( {0.0019,\;0.0170} \right)$, and the mean
square error is in the range of $\left( {3.2095e-13,\;2.2722e-9} \right)$.
We notice that Eq. (13) and Eq. (14) also satisfied here. From Figures 9-13,
the results of Gaussian noise fitting are the best, and the stretched
Gaussian noise fitting curves are better than the results of L\'{e}vy noise
data fitting.\\

\noindent{Table 6. }The estimated results for 1{\%} noise in the polynomial
case.
\begin{center}
\begin{tabular}{|l|p{66pt}|p{67pt}|p{71pt}|l|l|}
\hline
\textbf{\textit{function}}$_{\mathbf{}}$&
$a_{\mathbf{}}$&
$b_{\mathbf{}}$&
$c_{\mathbf{}}$&
$_{\mathbf{Rerr1}}$&
$_{\mathbf{Rerr2}}$ \\
\hline
$_{\mathbf{f}}$&
$_{4}$&
$_{3}$&
$_{2}$&
$_{0}$&
$_{0}$ \\
\hline
$_{\mathbf{FA}}$&
$_{3.8776}$&
$_{2.0197}$&
$_{2.9991}$&
$_{0.0019}$&
$_{3.2095e-13}$ \\
\hline
$_{\mathbf{FB}}$&
$_{3.9672\, \, \, \, \, \, \, }$&
$_{2.0072}$&
$_{3.0042}$&
$_{0.0046}$&
$_{4.1653e-10}$ \\
\hline
$_{\mathbf{FC}}$&
$_{3.8480\, \, \, \, \, \, \, }$&
$_{2.0262}$&
$_{2.9995}$&
$_{0.0013}$&
$_{7.4921e-14}$ \\
\hline
\end{tabular}
\label{tab6}
\end{center}
\begin{figure}[H]
\centering{\includegraphics[width=2in,height=1.6in]{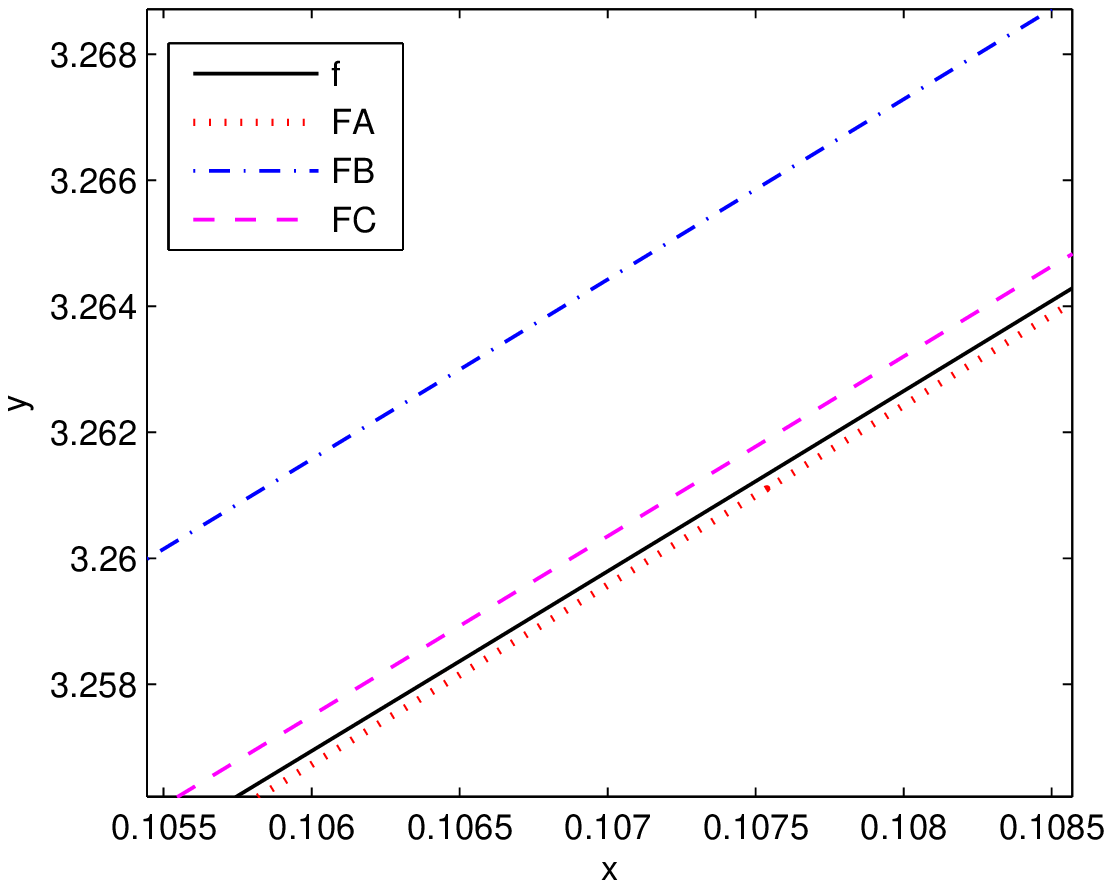}}

\textbf{Figure 9. Polynomial model fitting to 1{\%}noise.}
\label{fig9}
\end{figure}

\noindent{Table 7. }The estimated results for 5{\%} noise in the polynomial
case.
\begin{center}
\begin{tabular}{|l|p{65pt}|p{67pt}|p{71pt}|l|l|}
\hline
\textbf{\textit{function}}$_{\mathbf{}}$&
$a_{\mathbf{}}$&
$b_{\mathbf{}}$&
$c_{\mathbf{}}$&
$_{\mathbf{Rerr1}}$&
$_{\mathbf{Rerr2}}$ \\
\hline
$_{\mathbf{f}}$&
$_{4}$&
$_{3}$&
$_{2}$&
$_{0}$&
$_{0}$ \\
\hline
$_{\mathbf{FA}}$&
$_{3.7673\, \, \, \, \, \, \, \, }$&
$_{2.0500}$&
$_{2.9974}$&
$_{0.0026}$&
$_{1.1552e-12}$ \\
\hline
$_{\mathbf{FB}}$&
$_{4.6455\, \, \, \, \, \, \, \, \, \, \, \, }$&
$_{1.8825\, }$&
$_{3.0166}$&
$_{0.0190}$&
$_{3.4974e-8}$ \\
\hline
$_{\mathbf{FC}}$&
$_{4.7226\, \, \, \, \, \, \, \, \, \, \, \, \, }$&
$_{1.8644\, \, }$&
$_{2.9985}$&
$_{0.0079}$&
$_{1.1658e-9}$ \\
\hline
\end{tabular}
\label{tab7}
\end{center}
\begin{figure}[H]
\centering{\includegraphics[width=2in,height=1.6in]{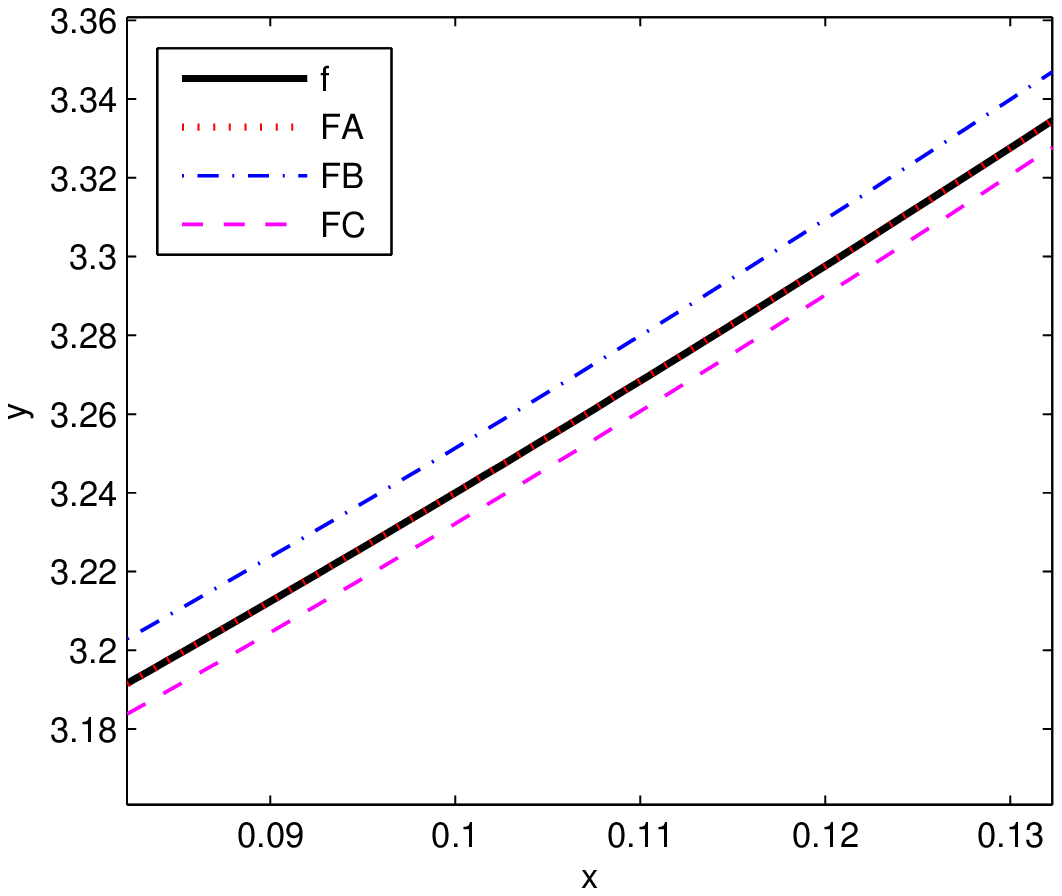}}

\textbf{Figure 10. Polynomial model fitting to 5{\%}noise.}
\label{fig10}
\end{figure}

\noindent{Table 8. }The estimated results for 10{\%} noise in the polynomial
case.
\begin{center}
\begin{tabular}{|l|p{64pt}|p{66pt}|p{70pt}|l|l|}
\hline
\textbf{\textit{function}}$_{\mathbf{}}$&
$a_{\mathbf{}}$&
$b_{\mathbf{}}$&
$c_{\mathbf{}}$&
$_{\mathbf{Rerr1}}$&
$_{\mathbf{Rerr2}}$ \\
\hline
$_{\mathbf{f}}$&
$_{4}$&
$_{3}$&
$_{2}$&
$_{0}$&
$_{0}$ \\
\hline
$_{\mathbf{FA}}$&
$_{4.4244}$&
$_{1.8693}$&
$_{3.0071}$&
$_{0.0071}$&
$_{7.7917e-11}$ \\
\hline
$_{\mathbf{FB}}$&
$_{4.6928}$&
$_{1.8345}$&
$_{3.0367}$&
$_{0.0367}$&
$_{7.6032e-7}$ \\
\hline
$_{\mathbf{FC}}$&
$_{3.5281}$&
$_{2.1336}$&
$_{2.9754}$&
$_{0.0246}$&
$_{9.9390e-8}$ \\
\hline
\end{tabular}
\label{tab8}
\end{center}
\begin{figure}[H]
\centering{\includegraphics[width=2in,height=1.6in]{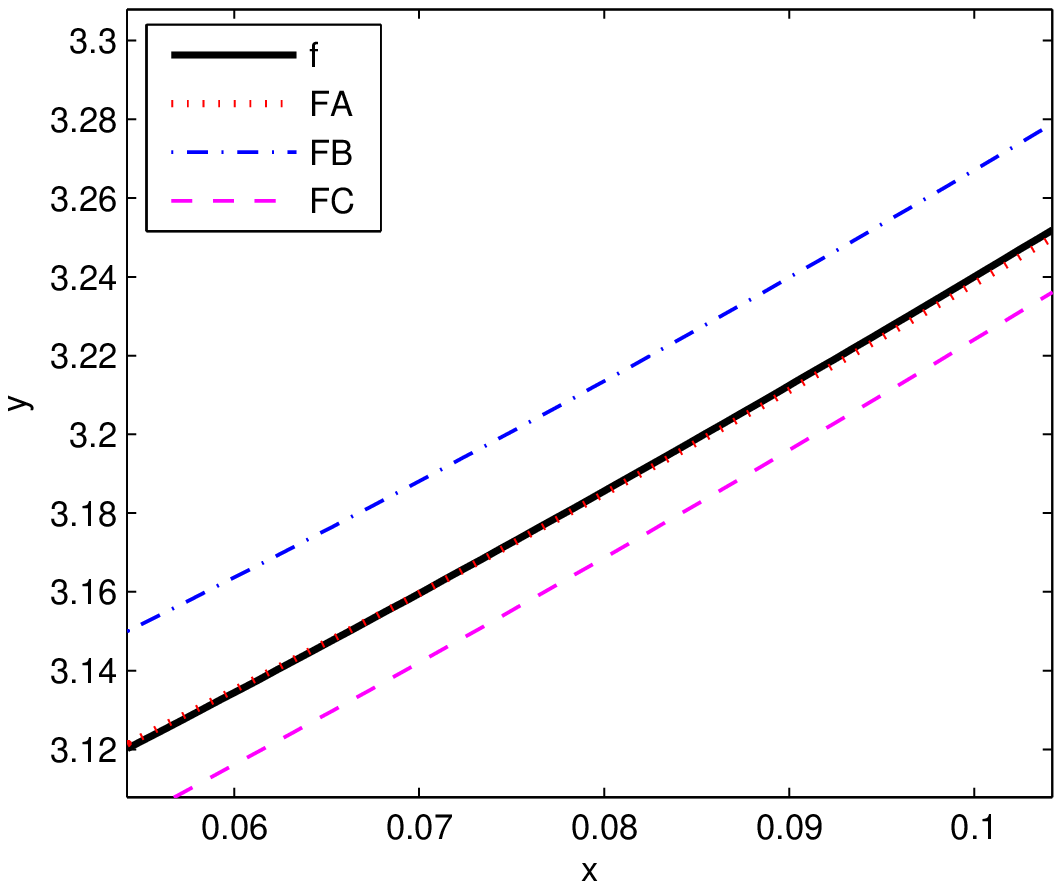}}

\textbf{Figure 11. Polynomial model fitting to 10{\%}noise.}
\label{fig11}
\end{figure}

\noindent{Table 9. }The estimated results for 15{\%} noise in the polynomial
case.
\begin{center}
\begin{tabular}{|l|p{66pt}|p{67pt}|p{71pt}|l|l|}
\hline
\textbf{\textit{function}}$_{\mathbf{}}$&
$a_{\mathbf{}}$&
$b_{\mathbf{}}$&
$c_{\mathbf{}}$&
$_{\mathbf{Rerr1}}$&
$_{\mathbf{Rerr2}}$ \\
\hline
$_{\mathbf{f}}$&
$_{4}$&
$_{3}$&
$_{2}$&
$_{0}$&
$_{0}$ \\
\hline
$_{\mathbf{FA}}$&
$_{4.9474}$&
$_{1.7866}$&
$_{3.0070}$&
$_{0.0128}$&
$_{5.1818e-10}$ \\
\hline
$_{\mathbf{FB}}$&
$_{4.3098}$&
$_{1.9391}$&
$_{3.0338}$&
$_{0.0379}$&
$_{1.1459e-6}$ \\
\hline
$_{\mathbf{FC}}$&
$_{4.4846\, \, \, \, \, \, \, \, \, \, \, \, \, \, \, \, \, \, \, \, \, \, \, }$&
$_{1.9551\, \, \, }$&
$_{2.9692\, }$&
$_{0.0318\, }$&
$_{5.2768e-7}$ \\
\hline
\end{tabular}
\label{tab9}
\end{center}
\begin{figure}[H]
\centering{\includegraphics[width=2in,height=1.6in]{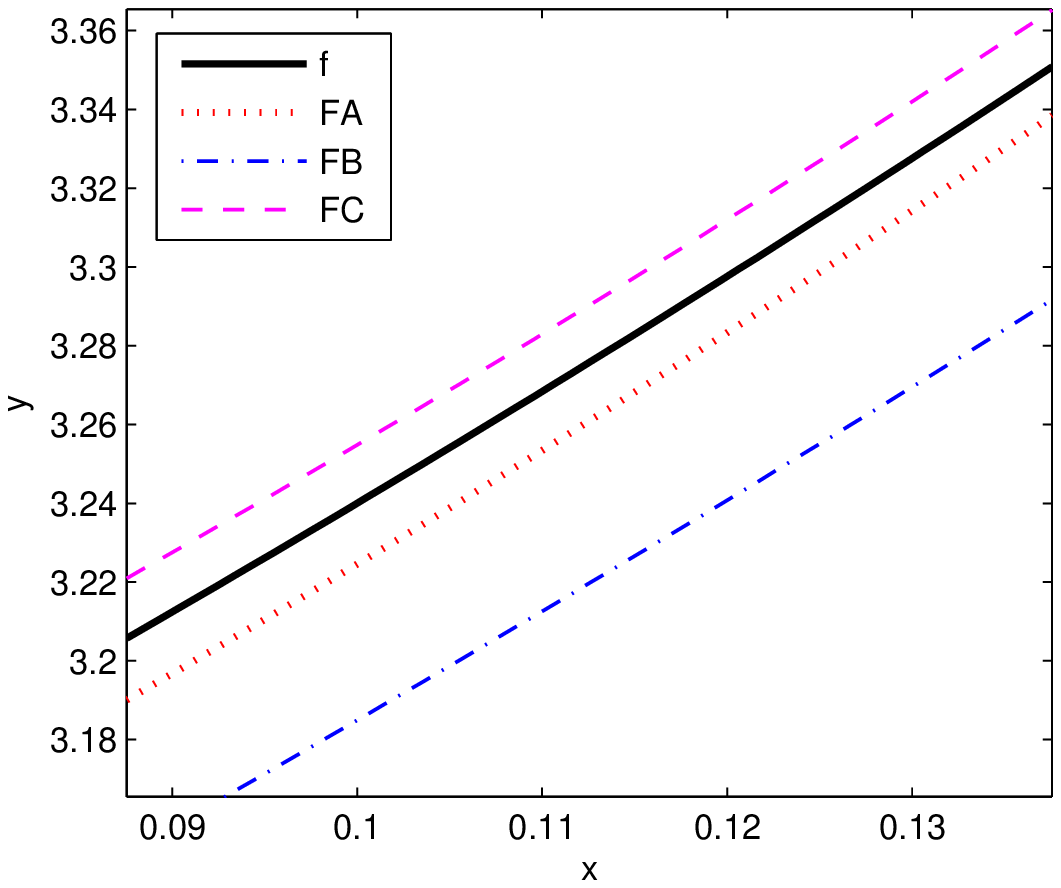}}

\textbf{Figure 12. Polynomial model fitting to 15{\%}noise.}
\label{fig12}
\end{figure}

\noindent{Table 10. }The estimated results for 20{\%} noise in the polynomial
case.
\begin{center}
\begin{tabular}{|l|p{65pt}|p{66pt}|p{70pt}|l|l|}
\hline
\textbf{\textit{function}}$_{\mathbf{}}$&
$a_{\mathbf{}}$&
$b_{\mathbf{}}$&
$c_{\mathbf{}}$&
$_{\mathbf{Rerr1}}$&
$_{\mathbf{Rerr2}}$ \\
\hline
$_{\mathbf{f}}$&
$_{4}$&
$_{3}$&
$_{2}$&
$_{0}$&
$_{0}$ \\
\hline
$_{\mathbf{FA}}$&
$_{6.1102\, \, \, \, \, \, \, \, \, \, \, \, \, \, \, \, \, \, \, \, \, \, \, \, }$&
$_{1.6197\, \, }$&
$_{3.0086\, \, \, }$&
$_{0.0170}$&
$_{2.2722e-9}$ \\
\hline
$_{\mathbf{FB}}$&
$_{3.0272\, \, \, \, \, \, \, \, \, \, \, \, \, \, \, \, \, \, \, \, \, \, \, \, \, \, \, \, \, }$&
$_{2.2146}$&
$_{3.0432\, \, \, }$&
$_{0.0550}$&
$_{7.1629e-6}$ \\
\hline
$_{\mathbf{FC}}$&
$_{4.0612\, \, \, \, \, \, \, \, \, \, \, \, \, \, \, \, \, \, \, \, \, \, \, \, \, \, \, }$&
$_{2.0446\, \, }$&
$_{2.9506\, \, \, \, }$&
$_{0.0494}$&
$_{3.8487e-6}$ \\
\hline
\end{tabular}
\label{tab10}
\end{center}
\begin{figure}[H]
\centering{\includegraphics[width=2in,height=1.6in]{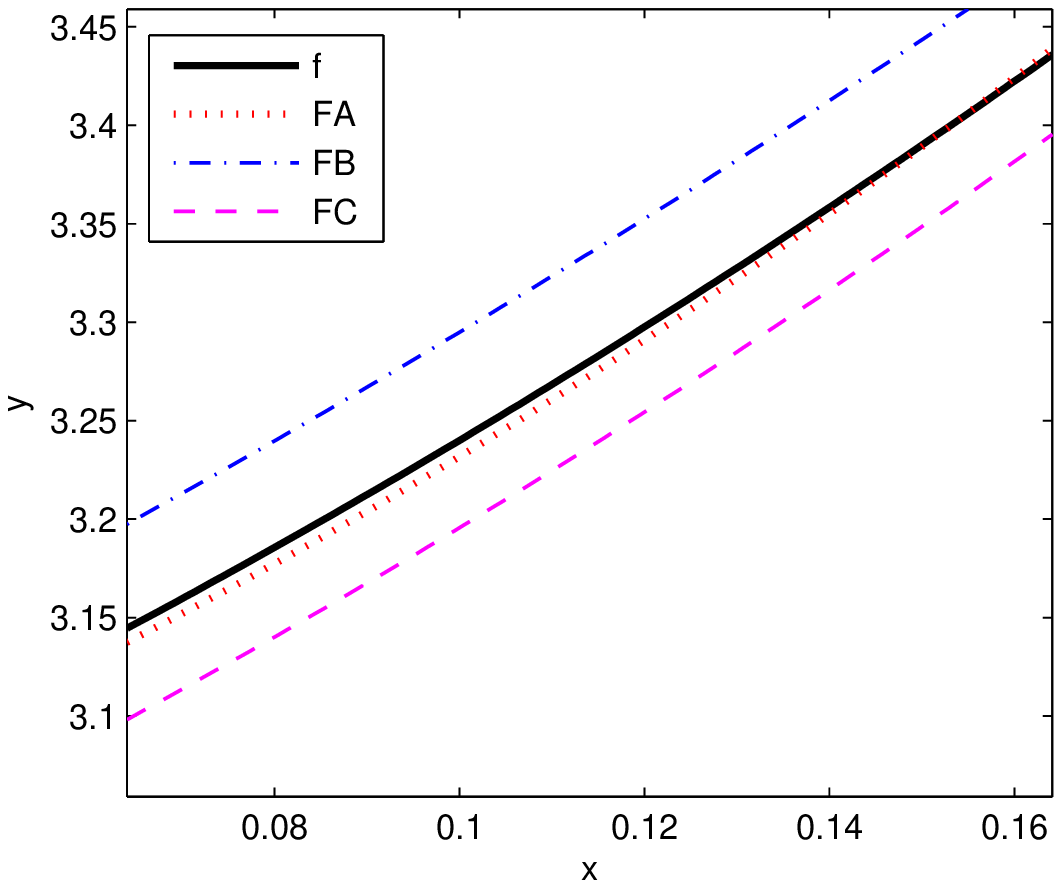}}

\textbf{Figure 13. Polynomial model fitting to 20{\%}noise.}
\label{fig13}
\end{figure}

c) Non-linear equations can be divided into two categories, one is
polynomial equation, and the other is non-polynomial equation. In this part,
we select the four parameters exponential function.
\begin{equation}
\label{eq18}
F\left( x \right)=ae^{3x}+be^{-t}-ct+d\;\left( {a\ne 0} \right).
\end{equation}

Here the following case is used as example:
\begin{equation}
\label{eq19}
f\left( x \right)=0.5e^{3x}+0.2e^{-t}-t+\frac{1}{3}.
\end{equation}

Tables 11 to 15 give the estimated parameters and the errors for five
different levels of noise in the exponential function case. The
corresponding fitting curves are shown in Figures 14-18. The Gaussian noise
fitting data maximum absolute error is in the range of $\left(
{0.0013,\;0.0091} \right)$, the mean square error is in the range of $\left(
{2.5659e-14,\;1.5559e-9} \right)$. And the results of exponential cases have
similar patterns with those shown in the linear and polynomial cases.\\

\noindent{Table 11. }The estimated results for 1{\%} noise in the exponential
case.
\begin{center}
\begin{tabular}{|l|p{54pt}|p{55pt}|p{57pt}|p{65pt}|l|l|}
\hline
\textbf{\textit{function}}$_{\mathbf{}}$&
$a_{\mathbf{}}$&
$b_{\mathbf{}}$&
$c_{\mathbf{}}$&
$d \quad_{\mathbf{}}$&
$_{\mathbf{Rerr1}}$ &
$_{\mathbf{Rerr2}}$ \\
\hline
$_{\mathbf{f}}$&
$_{0.5}$&
$_{0.2}$&
$_{1}$&
$_{1/3}$&
$_{0}$&
$_{0}$ \\
\hline
$_{\mathbf{FA}}$&
$_{0.4651\, \, \, \, \, \, \, \, \, }$&
$_{0.8096}$&
$_{0.3043}$&
$_{-0.2412}$&
$_{0.0013}$&
$_{2.5659e-14}$ \\
\hline
$_{\mathbf{FB}}$&
$_{0.4887\, \, \, \, \, \, \, \, \, \, \, \, }$&
$_{0.4365}$&
$_{0.7425}$&
$_{0.1168}$&
$_{0.0087}$&
$_{4.3599e-9}$ \\
\hline
$_{\mathbf{FC}}$&
$_{0.4889\, \, \, \, \, \, \, \, \, \, }$&
$_{0.3664}$&
$_{0.8068\, \, }$&
$_{0.1795}$&
$_{0.0015}$&
$_{7.7919e-13}$ \\
\hline
\end{tabular}
\label{tab11}
\end{center}
\begin{figure}[H]
\centering{\includegraphics[width=2in,height=1.6in]{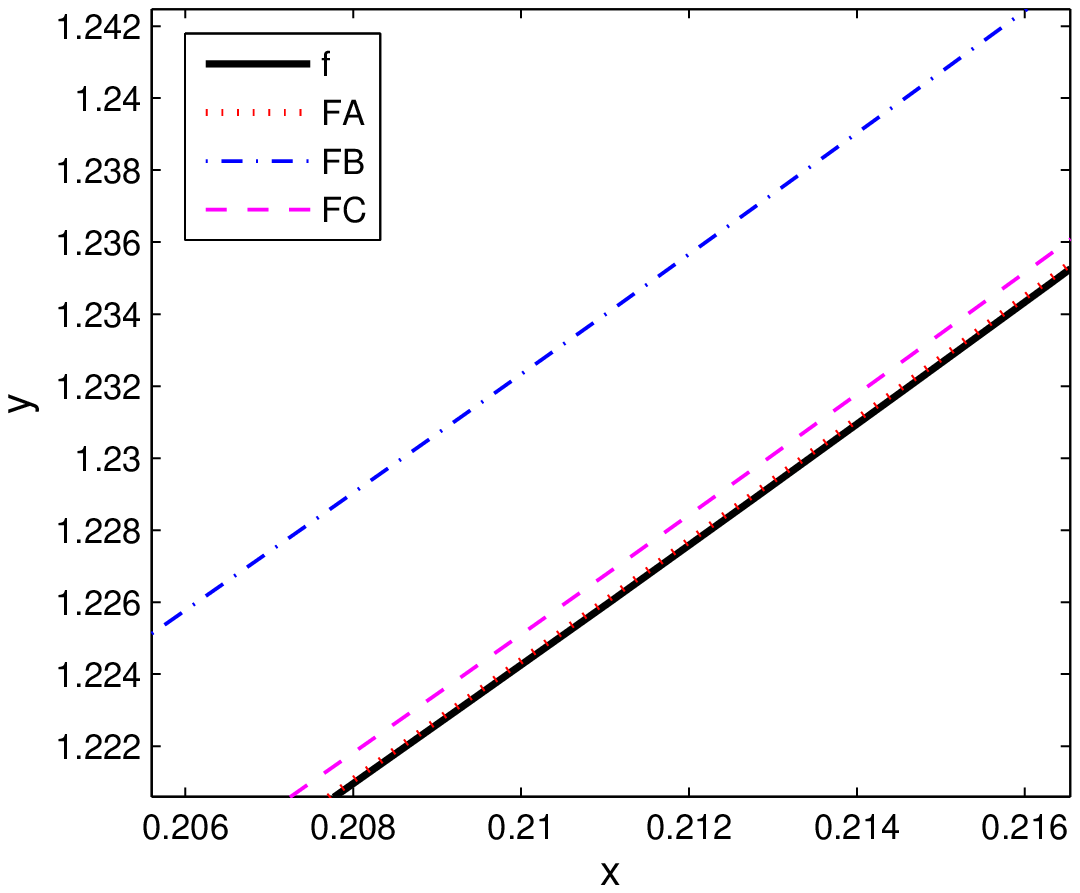}}

\textbf{Figure 14. Exponential model fitting to 1{\%} noise data.}
\label{fig14}
\end{figure}
\noindent{Table 12. }The estimated results for 5{\%} noise in the exponential
case.
\begin{center}
\begin{tabular}{|l|p{54pt}|p{54pt}|p{56pt}|p{64pt}|l|l|}
\hline
\textbf{\textit{function}}$_{\mathbf{}}$&
$a_{\mathbf{}}$&
$b_{\mathbf{}}$&
$c_{\mathbf{}}$&
$d \quad_{\mathbf{}}$&
$_{\mathbf{Rerr1}}$ &
$_{\mathbf{Rerr2}}$ \\
\hline
$_{\mathbf{f}}$&
$_{0.5}$&
$_{0.2}$&
$_{1}$&
$_{1/3}$&
$_{0}$&
$_{0}$ \\
\hline
$_{\mathbf{FA}}$&
$_{0.4282\, \, \, \, \, \, \, \, \, \, }$&
$_{1.5436}$&
$_{-0.5101}$&
$_{-0.9386}$&
$_{0.002}$&
$_{2.1618e-12}$ \\
\hline
$_{\mathbf{FB}}$&
$_{0.5493\, \, \, \, \, \, \, \, \, }$&
$_{-0.7494}$&
$_{2.0662}$&
$_{1.2535}$&
$_{0.0211}$&
$_{1.5321e-7}$ \\
\hline
$_{\mathbf{FC}}$&
$_{0.6557\, \, \, \, \, \, \, \, \, }$&
$_{-3.1130}$&
$_{4.5957}$&
$_{3.4751}$&
$_{0.0156}$&
$_{3.6588e-9}$ \\
\hline
\end{tabular}
\label{tab12}
\end{center}
\begin{figure}[H]
\centering{\includegraphics[width=2in,height=1.6in]{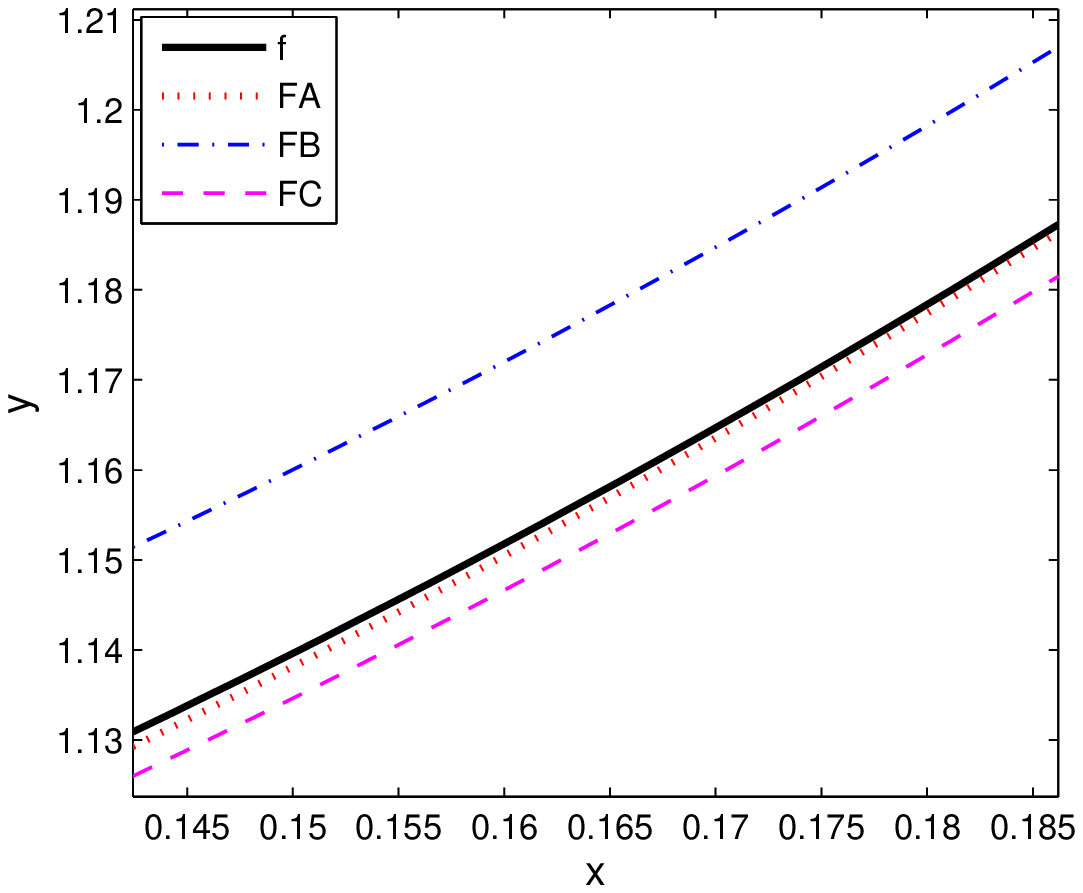}}

\textbf{Figure 15. Exponential model fitting to 5{\%} noise data.}
\label{fig15}
\end{figure}
\noindent{Table 13 }The estimated results for 10{\%} noise in the exponential
case.
\begin{center}
\begin{tabular}{|l|p{53pt}|p{54pt}|p{56pt}|p{64pt}|l|l|}
\hline
\textbf{\textit{function}}$_{\mathbf{}}$&
$a_{\mathbf{}}$&
$b_{\mathbf{}}$&
$c_{\mathbf{}}$&
$d \quad_{\mathbf{}}$&
$_{\mathbf{Rerr1}}$ &
$_{\mathbf{Rerr2}}$ \\
\hline
$_{\mathbf{f}}$&
$_{0.5}$&
$_{0.2}$&
$_{1}$&
$_{1/3}$&
$_{0}$&
$_{0}$ \\
\hline
$_{\mathbf{FA}}$&
$_{0.7448\, \, \, \, \, \, \, \, \, }$&
$_{-4.4825}$&
$_{6.2056}$&
$_{4.7692}$&
$_{0.0137}$&
$_{3.2450e-9}$ \\
\hline
$_{\mathbf{FB}}$&
$_{0.3697\, \, \, \, \, \, \, \, \, }$&
$_{2.9448}$&
$_{-1.9627}$&
$_{-2.3200}$&
$_{0.0545}$&
$_{6.0328e-6}$ \\
\hline
$_{\mathbf{FC}}$&
$_{0.8745\, \, \, \, \, \, \, \, \, }$&
$_{-6.9997}$&
$_{9.0002}$&
$_{7.1302}$&
$_{0.0284}$&
$_{6.2251e-8}$ \\
\hline
\end{tabular}
\label{tab13}
\end{center}
\begin{figure}[H]
\centering{\includegraphics[width=2in,height=1.5in]{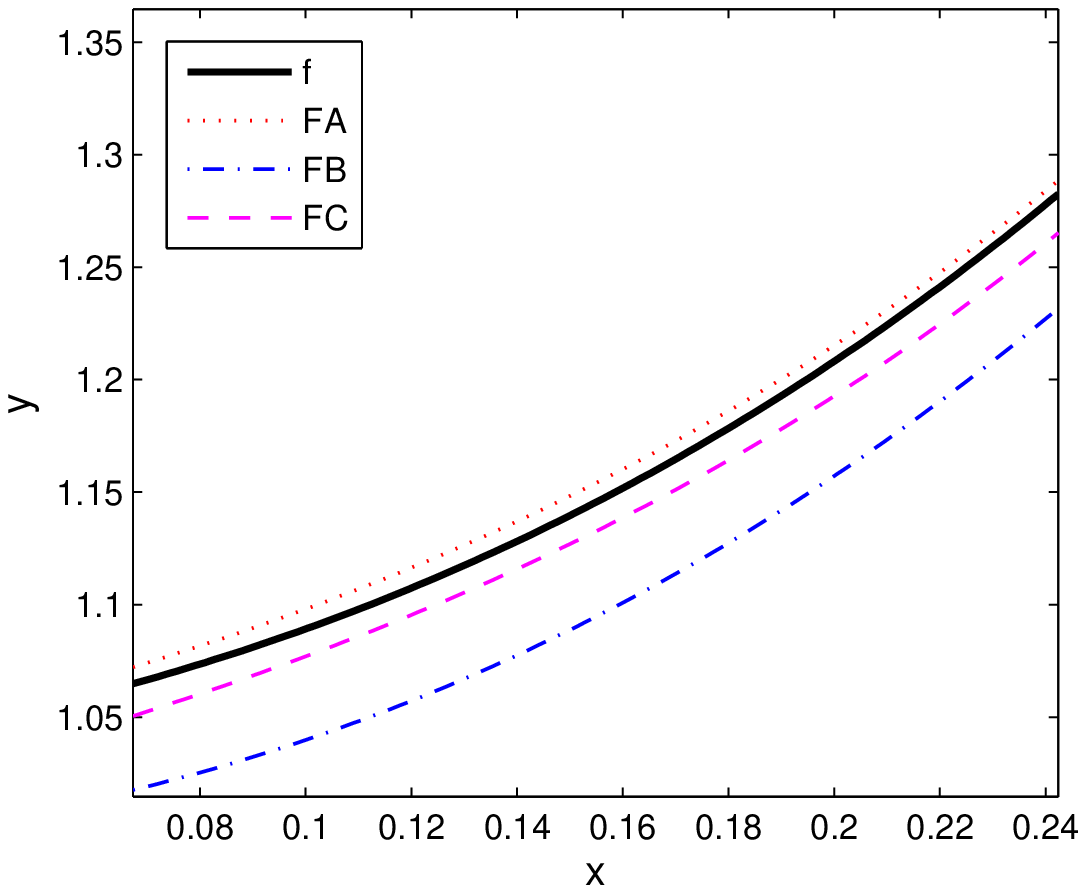}}

\textbf{Figure 16. Exponential model fitting to 10{\%} noise.}
\label{fig16}
\end{figure}

\noindent{Table 14. }The estimated results for 15{\%} noise in the exponential
case.
\begin{center}
\begin{tabular}{|l|p{55pt}|p{56pt}|p{58pt}|p{66pt}|l|l|}
\hline
\textbf{\textit{function}}$_{\mathbf{}}$&
$a_{\mathbf{}}$&
$b_{\mathbf{}}$&
$c_{\mathbf{}}$&
$d \quad_{\mathbf{}}$&
$_{\mathbf{Rerr1}}$ &
$_{\mathbf{Rerr2}}$ \\
\hline
$_{\mathbf{f}}$&
$_{0.5}$&
$_{0.2}$&
$_{1}$&
$_{1/3}$&
$_{0}$&
$_{0}$ \\
\hline
$_{\mathbf{FA}}$&
$_{0.1986\, \, \, \, \, \, \, \, \, }$&
$_{5.1211}$&
$_{-4.7226}$&
$_{-4.2859}$&
$_{0.0097}$&
$_{2.0389e-10}$ \\
\hline
$_{\mathbf{FB}}$&
$_{0.5151\, \, \, \, \, \, }$&
$_{-0.6695}$&
$_{1.8449\, \, \, \, }$&
$_{1.1120}$&
$_{0.0867}$&
$_{3.3841e-5}$ \\
\hline
$_{\mathbf{FC}}$&
$_{0.4250\, \, \, \, \, \, \, \, }$&
$_{1.1082}$&
$_{-0.1989}$&
$_{-0.5301}$&
$_{0.0303}$&
$_{1.8103e-7}$ \\
\hline
\end{tabular}
\label{tab14}
\end{center}
\begin{figure}[H]
\centering{\includegraphics[width=2in,height=1.6in]{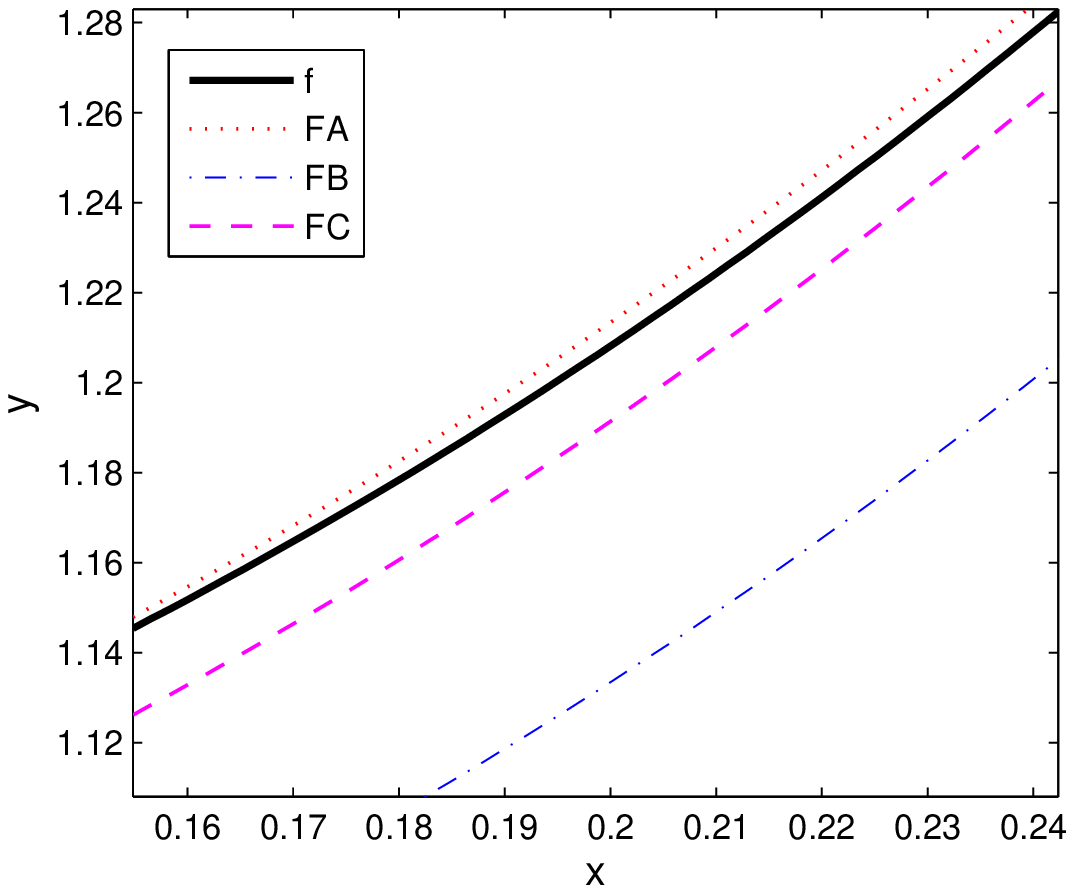}}

\textbf{Figure17. Exponential model fitting to 15{\%} noise.}
\label{fig17}
\end{figure}
\noindent{Table 15 }The estimated results for 20{\%} noise in the exponential
case.
\begin{center}
\begin{tabular}{|l|p{54pt}|p{55pt}|p{57pt}|p{65pt}|l|l|}
\hline
\textbf{\textit{function}}$_{\mathbf{}}$&
$a_{\mathbf{}}$&
$b_{\mathbf{}}$&
$c_{\mathbf{}}$&
$d \quad_{\mathbf{}}$&
$_{\mathbf{Rerr1}}$ &
$_{\mathbf{Rerr2}}$ \\
\hline
$_{\mathbf{f}}$&
$_{0.5}$&
$_{0.2}$&
$_{1}$&
$_{1/3}$&
$_{0}$&
$_{0}$ \\
\hline
$_{\mathbf{FA}}$&
$_{0.3164\, \, \, \, \, \, \, \, \, }$&
$_{3.9831}$&
$_{-3.1994}$&
$_{-3.2706}$&
$_{0.0091}$&
1.5559e-9 \\
\hline
$_{\mathbf{FB}}$&
$_{0.3339}$&
$_{3.7390}$&
$_{-2.8727}$&
$_{-2.9547}$&
$_{0.0873}$&
$_{4.5904e-5}$ \\
\hline
$_{\mathbf{FC}}$&
$_{0.6920}$&
$_{-3.3436}$&
$_{4.9695}$&
$_{3.6399}$&
$_{0.0451}$&
$_{1.8947e-6}$ \\
\hline
\end{tabular}
\label{tab15}
\end{center}
\begin{figure}[H]
\centering{\includegraphics[width=2in,height=1.6in]{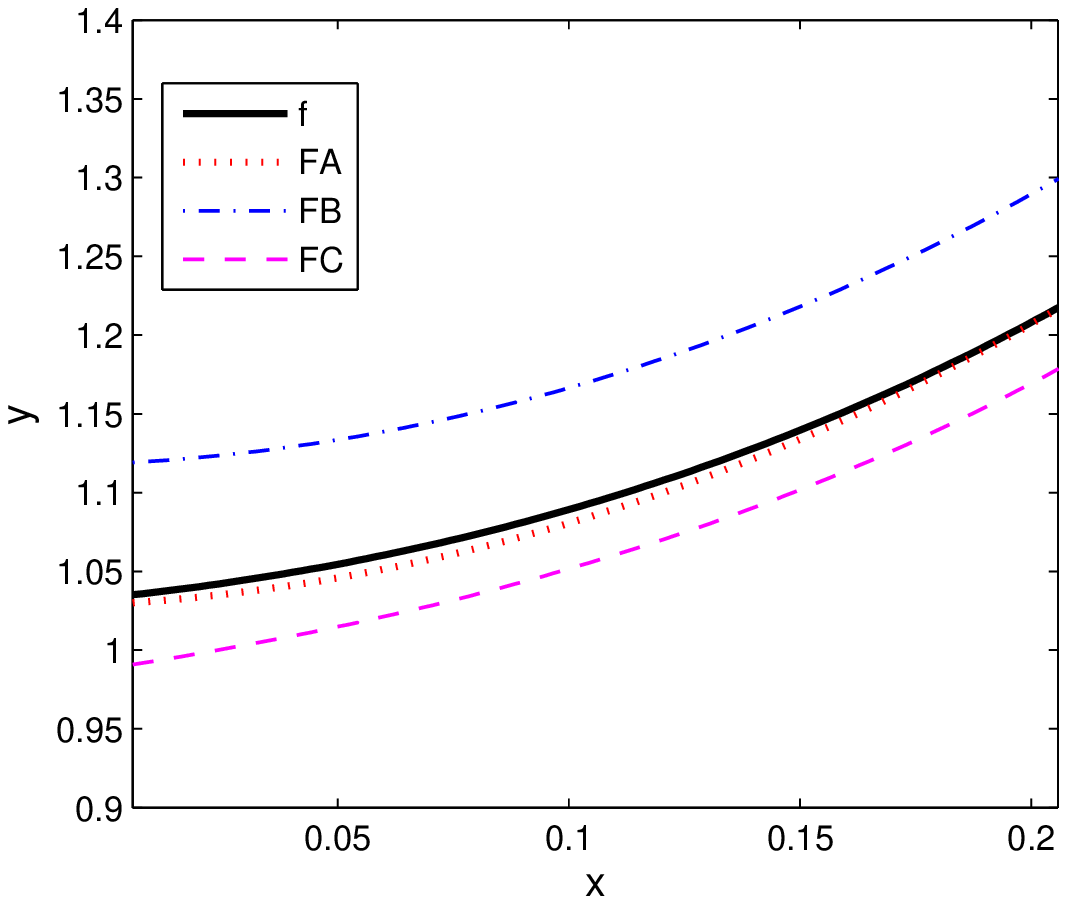}}

\textbf{Figure 18. Exponential model fitting to 20{\%} noise.}
\label{fig18}
\end{figure}

To summarize all the above results, we can find that the maximum absolute
and the mean square errors for the Gaussian noise cases are the smallest,
but the values for the L\'{e}vy noise cases are the biggest, i.e.,
\begin{equation}
\label{eq20}
Rerr1\left( {FA} \right)\mbox{\, }<\mbox{\, }Rerr1\left( {FC}
\right)\mbox{\, }<\mbox{\, }Rerr1\left( {FB} \right),
\end{equation}
\begin{equation}
\label{eq21}
Rerr2\left( {FA} \right)\mbox{\, }<\mbox{\, }Rerr2\left( {FC}
\right)\mbox{\, }<\mbox{\, }Rerr2\left( {FB} \right).
\end{equation}

It can be observed from Figures 4 to 18, that the results of Gaussian noise
fitting have the best accuracy, and the stretched Gaussian noise fitting
curves are closer to those of the Gaussian noise compared with the results
of L\'{e}vy noise data fitting. Thus, the least square method is less
accurate when it is applied to the non-Gaussian noise data fitting compared
with the cases of Gaussian noise, especially when the noise level is larger
than 5{\%}.

This study mainly verifies the least square method is inapplicable to
non-Gaussian noise when the noise level is high. To extend the results in
more complicated systems, a mathematical proof to the conclusion should be
derived in future study.

The second goal of our further work is to modify the least square method in
fitting non-Gaussian noises. Actually the core concept of the least square
method is to minimize the square error $\delta =\sum\limits_{i=1}^n {\left[
{y_{i} -f\left( {x_{i} } \right)} \right]^{2}} $, i.e., in the linear case,
\begin{equation}
\label{eq22}
\delta =\sum\limits_{i=1}^n {\left[ {yi-ax-b} \right]^{2}} ,
\end{equation}
to compute the minimum value of Eq. (22), the main task is to set the
first-order derivatives of the parameters to be zero.
\begin{equation}
\label{eq23}
\left\{ {\begin{array}{l}
 \frac{\partial \delta }{\partial a}=0 \\
 \frac{\partial \delta }{\partial b}=0 \\
 \end{array}} \right.\quad .
\end{equation}

The solutions of Eq. (23) are the target values of the parameters $a$ and $b$.
Combining our previous work on fractional and fractal derivatives,$^{\,
30,36}$ we can employ the fractional and fractal derivatives to generalize
Eq. (23), and the corresponding fitting errors can be defined by using the
following power law transform:
\begin{equation}
\label{eq24}
\hat{{x}}=x^{\beta }.
\end{equation}\\

\noindent\textbf{4. Conclusions }\\

This study examines the feasibility of least square method in fitting
various noise data polluted by adding different levels of Gaussian and
non-Gaussian noise to exact values of the selected functions including
linear equations, polynomial and exponential equations. The maximum absolute
error and the mean square error are calculated and compared for the
different cases. Based on the foregoing results and discussions, the
following conclusions can be drawn:

1. The fitting results for the non-Gaussian noise are less accurate than
those of the Gaussian noise, but the stretched Gaussian cases appear to
perform better than the L\'{e}vy noise cases.

2. The least-squares method is inapplicable to the non-Gaussian noise data
when the noise level is larger than 5{\%}.

3. A theoretical proof and improved least mean square methods for
non-Gaussian noise data are under intense study.\\

\noindent\textbf{Acknowledgments}\\

This paper was supported by the National Science Funds for
Distinguished Young Scholars of China (Grant No. 11125208) and the
111 project (Grant No. B12032).\\

\noindent\textbf{References}\\

1. D. Middleton. Non-Gaussian noise models in signal processing for
telecommunications: new methods an results for class A and class B
noise models. \textit{IEEE Transactions on Information Theory} 1999;
45(4): 1129-1149.

2. A. Nasri, R. Schober, Y. Ma. Unified asymptotic analysis of linearly
modulated signals in fading, non-Gaussian noise and interference.
\textit{Communications IEEE Transactions on} 2008; 56(6): 980-990.

3. X. Wang, R. Chen. Blind turbo equalization in Gaussian and impulsive
noise. \textit{IEEE Transactions on Vehicular Technology~}2001; 50(4):1092-1105.

4. R. Blum, R. Kozick, B. Sadler. An adaptive spatial diversity receiver for
non--Gaussian interference and noise. \textit{IEEE Trans. Signal Processing }1999; 47: 2100-2111.

5. L. He, Y. Cui, T. Zhang. Analysis of weak signal detection based on
tri-stable system under Levy noise. \textit{Chinese Physics B: English }2016; 6: 85-94.

6. Y. Zhao, X. Zhuang, S. J. Ting. Gaussian mixture density modeling of
non-Gaussian source for autoregressive process. \textit{IEEE Transactions on Signal Processing }1995; 43(4): 894-903.

7. S. Chen, B. Mulgrew, L. Hanzo. Least bit error rate adaptive nonlinear
equalizers for binary signaling. \textit{IEEE Proceedings Communications }2003; 150(1): 29-36.

8. C.H. Chapman, J. A. Orcutt. Least-square fitting of marine seismic
refraction data. \textit{Geophysical Journal International 1985};~82(3): 339-374.

9. A. Nasri, R. Schober. Performance of BICM-SC and BICM-OFDM systems with
diversity reception in non-Gaussian noise and interference. \textit{IEEE Transactions on Communications~}2009;~57(11):
3316-3327.

10. A. Aldo Faisal, Luc P. J. Selen,~Daniel M. Wolpert. Noise in the nervous
system. \textit{Nature Reviews Neuroscience}~2008,~9(4): 292-303.

11. E. Dobierzewska-Mozrzymas, G. Szymczak, P. Biega\'{n}ski, E. Pieciul.
L\'{e}vy's distributions for statistical description of fractal structures;
discontinuous metal films on dielectric substrates. Physica B Condensed
Matter 2003; 337(1--4): 79--86.

12. F. Ren, Y. Xu, W. Qiu, J. Liang. Universality of stretched Gaussian
asymptotic diffusion behavior on biased heterogeneous fractal structure in
external force fields. \textit{Chaos Solitons {\&} Fractals }2005; 24(1):273-278.

13. T. Solomon, E. Weeks, H. Swinney. Observation of anomalous diffusion and
L\'{e}vy flights in a two-dimensional rotating flow.\textit{ Physical Review Letters }1993;~71(24): 3975-3978.

14. L. Chrysostomos. Signal processing with alpha-stable distributions and
applications. J\textit{ohn Wiley {\&} Sons, Inc. }1995; 22(3): 333-334.

15. R. Gomory, B. Mandelbrot. Fractals and scaling in finance:
discontinuity, concentration, Risk. New York: Springer, 1997.

16. I. Sokolov, W. Ebelling, B. Dybiec, Harmonic oscillator under L\'{e}vy
noise: Unexpected properties in the phase space. \textit{Phys. Rev. E. }2011; 83 (041118).

17. R. Segev, M. Benveniste, E. Hulata, et al. Long term behavior of
lithographically prepared in vitro neuronal networks. \textit{Phys. Rev. Lett}. 2002; 88: 11-18.

18. H. Long, C. Ma, Y. Shimizu. Least squares estimators for stochastic
differential equations driven by small L\'{e}vy noises. \textit{Stochastic Processes {\&} Their Applications} 2016; 8(6): 1-21.

19. W.H. Liao, A. Roebel, W.Y. Su. On stretching Gaussian noises with the
phase vocoder. \textit{Proc Int Conf on Digital Audio Effects}~2012; 9(15): 17-21.

20. L. G. Alves, D. B. Scariot, R. R. Guimar\~{a}es, et al.. Transient
superdiffusion and Long-Range correlations in the Motility patterns of
trypanosomatid flagellate protozoa. \textit{Plos One }2016; 11(3): e0152092.

21. S. Sugita, M. Yagi, S. Itoh, K. Itoh. Bohm-like dependence of transport
in scrape-off layer plasmas. \textit{Journal of the Physical Society of Japan} 2012; 81(4): 69-69.

22. F. Ren, J. Wang, L. Lv, H. Pan, W. Qiu. Effect of different waiting time
processes with memory to anomalous diffusion dynamics in an external force
fields. \textit{Contents lists available at Science Direct: Physical A} 2015; 417: 202--214.

23. N. N. Kolchigin, S. N. Pivnenko. Numerical modeling of measurements of
dielectric material characteristics using non-sinusoidal signals. \textit{Proc Int Conf on Digital Audio Effects}~2012; 9:
17-21.

24. P. Lancaster, K. \v{S}alkauskas,~Curve and Surface Fitting: An
Introduction.~London: Academic Press, 1986.

25. D. N. Lehmer. Review: E. T. Whittaker and G. Robinson, the calculus of
observations. A treatise on numerical mathematics. \textit{Phys. Rev. Lett}. 1925; 98(6):
068102-068102.

26. X. Chen. Concise history of statistics. Changsha: Hunan Education
Publishing House, 2002.

27. H. Xian. Study on ANN Noise Adaptability in Application of Industry
Process Characteristics Mining.\textit{ Advanced Materials Research} 2012; 462:635-640.

28. G. A. Tsihrintzis, C. L.Nikias. Fast estimation of the parameters of
alpha-stable impulsive interference. \textit{IEEE Transactions on Signal Processing} 1996; 44(6): 1492-1503.

29. W. Chen, H. Sun, X. Li .The fractional derivative model of mechanics and
engineering problems. Beijing: Science Press, 2010.

30. W Chen, H Sun, X Zhang, D Koro�ak. Anomalous diffusion modeling by
fractal and fractional derivatives. \textit{Computers {\&} Mathematics with Applications} 2010;~59(5): 1754-1758.

31. W. Chen, Time-space fabric underlying anomalous diffusion. \textit{Chaos Solitons {\&} Fractals} 2005; 28(4):
923-929.

32.B. Wang, Y. Wei, Y. Zhang. Generate random number by using Inverse
function and transform sampling method. \textit{Journal of Ningxia Teachers University} ~2012; 33(3): 24-28.

33. B. Wang, Y. Wei, Y. Sun. Generate random number by using acceptance
rejection method. \textit{Journal of Chongqing normal university (Natural science)} 2013; 30(6): 86-91.

34. J. M. Chambers, C. L. Mallows, B. W. Stuck. A method for simulating
stable random variables. J\textit{ournal of the American Statistical Association} 1976; 71(354): 340-344.

35. R.Weron. Computationally intensive value at risk calculations. \textit{Hand book of Computational Statistics}. Berlin:
Springer, 2004.

36. W. Chen, L. Ye, H. Su. Fractional diffusion equations by the
Kansa method. \textit{Computers and Mathematics with Applications}
2010; 59: 1614--1620.

\end{document}